\documentclass[trackchanges]{aastex701}

\usepackage{bm}
\usepackage{amsfonts}
\usepackage{latexsym}
\usepackage{graphicx}
\usepackage{amsmath}
\usepackage{palatino}
\usepackage{mathpazo}
\usepackage{textcomp}
\usepackage{float}
\usepackage{booktabs}
\usepackage{dcolumn}
\usepackage{multirow}
\usepackage{array} 
\usepackage{tabularx}

\usepackage{ragged2e}
\usepackage{tensor}

\usepackage[version=3]{mhchem}

\usepackage{xcolor}
\usepackage{enumitem}
\usepackage{orcidlink}
\usepackage{epsfig}
\usepackage{commath}
\usepackage{caption}
\usepackage{subcaption}

\usepackage[latin1]{inputenc}
\usepackage{hyperref}

\begin{document}

\title{Low-mass and ultra-massive white dwarfs as separate probes of trace-coupled teleparallel gravity}

\author[orcid=0000-0002-9265-5025, gname=Rajasmita,sname=Sahoo]{Rajasmita Sahoo}
\affiliation{National Institute of Technology, Tiruchirappalli, Tamil Nadu, 620015, India.}
\email[show]{rsphysics58@gmail.com}

\author[orcid=0000-0003-2082-697X,gname={Devaraja Mallesha},sname= Naik]{Devaraja Mallesha Naik}
\affiliation{Department of Mathematics, Government First Grade College Shikaripura, 
Shivamogga--577 427, Karnataka, India.}
\email[show]{devarajamaths@gmail.com}  
\correspondingauthor{Devaraja Mallesha Naik}






\begin{abstract}
We study the equilibrium structure of white dwarfs in $f(T,\mathcal{T})$ gravity, in which the torsion scalar $T$ is coupled to the trace $\mathcal{T}$ of the energy--momentum tensor. We consider the model $f(T,\mathcal{T}) = T + \alpha T^2 + \beta\mathcal{T}$ with the rotated tetrad and derive the stellar structure equations in a single, consistent set of conventions. The trace coupling makes matter non-conserved, and the hydrostatic equation acquires a term proportional to $\beta/c_s^2$. Since the sound speed is very small in the outer layers, regular stars require $\beta\le 0$. The quadratic term survives outside the star, so the gravitational mass must be taken at infinity. The two single-coupling limits of the model coincide with $f(R,\mathcal{T})$ and quadratic $f(T)$ gravity, and our solver reproduces earlier white-dwarf results in both. With the Chandrasekhar equation of state, a positive $\alpha$ lowers the maximum mass from $1.416\,M_\odot$ in general relativity to $1.285\,M_\odot$ at $\alpha=10^{20}\,\mathrm{cm^2}$. For $\alpha<0$ there is no turning point, and the mass is limited only by electron capture. A negative $\beta$ acts mainly on low-mass white dwarfs, whose outer layers it inflates. The two couplings therefore probe different parts of the mass--radius relation. Using Sirius~B, 40~Eri~B and Procyon~B, whose masses are known dynamically, we find $|\beta|\lesssim(2$--$4)\times10^{-5}$, independently of $\alpha$  for $-10^{18}\le\alpha\le10^{20}\,\mathrm{cm^2}$. A conservative bound that discards layers below $10^6\,\mathrm{g\,cm^{-3}}$ is $|\beta|\lesssim(1$--$2)\times10^{-4}$. The ultra-massive white dwarf ZTF~J1901+1458 suggests $\alpha\lesssim3\times10^{19}\,\mathrm{cm^2}$. Combinations of $\alpha$ and $\beta$ that mimic the general-relativistic maximum mass are excluded by the radii of low-mass white dwarfs.
\end{abstract}


\section{Introduction}
\label{sec:intro}

General relativity (GR) has passed every precision test to date. These range from the Solar System and binary pulsars \citep{Will2014} to the direct detection of gravitational waves \citep{Abbott2016,Abbott2017} and the image of a black-hole shadow \citep{EHT2019}. On cosmological scales, however, the accelerated expansion of the Universe \citep{Riess1998,Perlmutter1999} requires either a dark-energy component or a modification of gravity. The simplest option, a cosmological constant, faces a severe fine-tuning problem \citep{Weinberg1989}. Tensions between early- and late-Universe measurements, most notably in the Hubble constant, add further motivation \citep{Abdalla2022}. These open problems have motivated a large class of modified gravity theories \citep{DeFeliceTsujikawa2010,CapozzielloDeLaurentis2011,Clifton2012,Nojiri2017}.
Most of these theories are built to reproduce GR in the Solar System and to modify gravity on cosmological scales. Compact stars probe them in a third regime, that of high densities and strong gravitational fields. In this regime the coupling between matter and geometry can differ from GR, while the matter itself is relatively well understood. Compact stars are therefore natural testing grounds for modified gravity \citep{OlmoRubieraWojnar2020}.

Teleparallel gravity offers an alternative geometric description of gravitation. Its fundamental field is the tetrad, and gravity is carried by the torsion of the curvature-free Weitzenb\"ock connection rather than by curvature \citep{AldrovandiPereira2013}. The teleparallel equivalent of general relativity (TEGR) uses the torsion scalar $T$ as its Lagrangian. It differs from the Einstein--Hilbert action only by a boundary term and has the same field equations as GR. Promoting $T$ to a function $f(T)$ gives a modified theory whose field equations remain of second order, unlike those of $f(R)$ gravity \citep{FerraroFiorini2007,BengocheaFerraro2009,CaiReview2016,Bahamonde2023}. In nonlinear $f(T)$ the choice of tetrad becomes physical \citep{LiSotiriouBarrow2011}. This is handled either by working with a ``good'' tetrad \citep{TamaniniBoehmer2012} or by including the inertial spin connection in a covariant formulation \citep{KrssakSaridakis2016,Krssak2019,Golovnev2021}.

Relativistic stars exist in $f(T)$ gravity \citep{BohmerMussaTamanini2011}. Stellar models in the quadratic model $f(T)=T+\alpha T^2$ have been built for polytropes \citep{IlijicSossich2018} and for realistic equations of state \citep{LinChenZhai2022,deAraujoFortes2023}. Power-law extensions have also been studied \citep{deAraujoFortes2023b}. A subtle point is the definition of the mass. For $f_{TT}\neq0$ and the tetrads used in stellar models, the Schwarzschild metric is not a vacuum solution, and the exterior is known only perturbatively \citep{DeBenedictisIlijic2016}. The asymptotic mass measured at infinity has therefore been advocated as the appropriate definition \citep{deAraujoFortes2023}. White dwarfs have recently been studied in quadratic $f(T)$ gravity with this definition and with a Coulomb-corrected equation of state \citep{Otoniel2026}, and in exponential $f(T)$ gravity with envelope and temperature corrections \citep{Farrugia2025}.
A further extension couples the torsion scalar to the trace $\mathcal{T}$ of the matter energy--momentum tensor. This $f(T,\mathcal{T})$ gravity \citep{Harko2014} is the teleparallel counterpart of $f(R,\mathcal{T})$ gravity \citep{HarkoFRT2011}. The trace coupling makes the theory non-conservative, and matter is not separately conserved. Compact stars in $f(T,\mathcal{T})$ gravity have so far been studied as quark stars \citep{PaceSaid2017a,Tangphati2026} and neutron stars \citep{PaceSaid2017b}. As we discuss in Sec.~\ref{sec:framework}, these studies carried the matter coupling of the original formulation into the opposite metric signature, so their effective sources do not correspond to the matter Lagrangian they adopt.

White dwarfs deserve particular attention among compact stars. They are the end point of most stars, and their structure is set by electron degeneracy pressure. In GR this leads to the Chandrasekhar limit of about $1.4\,M_\odot$ \citep{Chandrasekhar1931,ShapiroTeukolsky1983}. Their microphysics is well understood, and several white dwarfs have precise masses and radii \citep{Saumon2022,Bond2017Sirius,Bond2017EriB,Bond2015Procyon}. Observations have also raised questions about the universality of this limit. Overluminous type Ia supernovae have been interpreted in terms of super-Chandrasekhar progenitors \citep{Howell2006,Hicken2007,Scalzo2010}, and underluminous ones in terms of sub-Chandrasekhar progenitors \citep{Filippenko1992,Taubenberger2008}. Ultra-massive white dwarfs close to the limit have also been found \citep{Caiazzo2021}.
These developments have made white dwarfs a popular laboratory for modified gravity. They have been studied in $f(R)$ gravity \citep{DasMukhopadhyay2015,Kalita:2018ldn,KalitaSarmah2022}, Palatini $f(R)$ gravity \citep{Sarmah2022,Kalita2023}, scalar--tensor theories beyond Horndeski \citep{Jain2016,Saltas2018}, massive scalar--tensor theories \citep{Vidal2025}, fourth-order and other theories \citep{Banerjee2017}, massive gravity \citep{EslamPanahLiu2019} and Rastall--Rainbow gravity \citep{Li2024}. Theories with matter--geometry couplings have received particular attention: $f(R,\mathcal{T})$ gravity \citep{Carvalho2017,Rocha2020}, $f(R,L_m)$ gravity \citep{Lobato2022}, $f(R,L_m,\mathcal{T})$ gravity \citep{OtonielFRLT2026,Priyobarta2026}, and $f(Q)$ gravity \citep{Sahoo2026,OtonielFQ2026}. In trace-coupled theories the hydrostatic equation contains the inverse sound speed \citep{Moraes2016,Carvalho2017}. Degenerate electron matter has a very small sound speed, which makes white dwarfs especially sensitive to such couplings.

Two aspects of this literature motivate the present work. First, many studies aim at super-Chandrasekhar masses. In some cases these are reached only at central densities where electron captures would already have changed the composition \citep{Priyobarta2026}. Second, observational data are often used only for qualitative comparison, rather than to constrain the couplings. White dwarfs with dynamical masses are particularly well suited for quantitative tests, because their masses do not depend on an assumed mass--radius relation \citep{Bond2017Sirius,Bond2017EriB,Bond2015Procyon}.

In this work we study white dwarfs in $f(T,\mathcal{T})$ gravity for the model $f=T+\alpha T^2+\beta\mathcal{T}$. To our knowledge, this is the first study of white dwarfs in this theory. Its two single-coupling limits are known theories. For $\alpha=0$ the model is equivalent to $f(R,\mathcal{T})=R+2\lambda\mathcal{T}$ \citep{HarkoFRT2011} with $\lambda=\beta/2$ \citep{Carvalho2017}. For $\beta=0$ it is the quadratic $f(T)$ model of \citep{Otoniel2026}. We use both limits to validate our solver, and we focus on what is new: the combined model. We impose the electron-capture density as an upper limit, and we use the mass measured at infinity throughout. We show that the two couplings act on different parts of the mass--radius relation. The quadratic term acts only near the maximum mass, while the trace coupling acts mostly on low-mass stars. As a result, the bound on $\beta$ from white dwarfs with dynamical masses is independent of $\alpha$ for $-10^{18}\le\alpha\le10^{20}\,\mathrm{cm^2}$. Through the equivalence with $f(R,\mathcal{T})$ gravity, this bound also improves the corresponding white-dwarf limit by more than an order of magnitude, or by a factor of 3--6 if only the degenerate interior is used. We also give a simple estimate of the difference between the surface mass and the mass measured at infinity, $M_\infty/m(R)-1\simeq2\alpha T(R)$, which follows from the vacuum solution of \citep{DeBenedictisIlijic2016}. Finally, we examine how robust the bound on $\beta$ is to the treatment of the outer layers.

The paper is organized as follows. Section~\ref{sec:framework} presents the theory and fixes the conventions. Section~\ref{sec:structure} derives the stellar structure equations, the exterior mass correction and the equation of state. Section~\ref{sec:results} presents the validation, the stellar structure, the mass--radius relations, the stability and energy conditions, and the observational constraints. Section~\ref{sec:conclusions} summarizes our findings. Two appendices describe the Coulomb-corrected equation of state, used to compare with quadratic $f(T)$ gravity, and the role of the surface definition.

\section{Teleparallel gravity and $f(T,\mathcal{T})$ theories}
\label{sec:framework}

In teleparallel gravity the fundamental dynamical variable is the tetrad $e^{a}{}_{\mu}$,  from which the metric is derived as
\begin{equation}
g_{\mu\nu} = \eta_{ab}\,e^{a}{}_{\mu}e^{b}{}_{\nu}, \qquad e\equiv\det(e^{a}{}_{\mu})=\sqrt{-g},
\label{eq:metric}
\end{equation}
where Greek indices label spacetime coordinates, Latin indices label the tangent space, and $\eta_{ab}=\mathrm{diag}(-1,1,1,1)$. The inverse tetrad $e_{a}{}^{\mu}$ satisfies $e_{a}{}^{\mu}e^{a}{}_{\nu}=\delta^{\mu}_{\nu}$ and $e_{a}{}^{\mu}e^{b}{}_{\mu}=\delta^{b}_{a}$. Throughout this work we use the signature $(-,+,+,+)$ and  set $G=c=1$; physical units are restored in Sec.~\ref{sec:structure} for the numerical solutions. Gravity is described by the curvature-free Weitzenb\"ock connection $\Gamma^{\lambda}{}_{\nu\mu}=e_{a}{}^{\lambda}\partial_{\mu}e^{a}{}_{\nu}$, whose antisymmetric part defines the torsion, $T^{\lambda}{}_{\mu\nu}=\Gamma^{\lambda}{}_{\nu\mu}-\Gamma^{\lambda}{}_{\mu\nu}$, so that all gravitational information is carried by torsion. In the pure-tetrad formulation, where the spin connection vanishes, the torsion tensor reads
\begin{equation}
T^{\lambda}{}_{\mu\nu} = e_{a}{}^{\lambda}\left(\partial_\mu e^{a}{}_{\nu}-\partial_\nu e^{a}{}_{\mu}\right),
\label{eq:torsion}
\end{equation}
and from it one constructs the contortion and the superpotential,
\begin{equation}
K^{\mu\nu}{}_{\rho} = -\tfrac{1}{2}\left(T^{\mu\nu}{}_{\rho}-T^{\nu\mu}{}_{\rho}-T_{\rho}{}^{\mu\nu}\right), \qquad
S_{\rho}{}^{\mu\nu} = \tfrac{1}{2}\left(K^{\mu\nu}{}_{\rho}+\delta^{\mu}_{\rho}T^{\alpha\nu}{}_{\alpha}-\delta^{\nu}_{\rho}T^{\alpha\mu}{}_{\alpha}\right).
\label{eq:KS}
\end{equation}
The contraction of the superpotential with the torsion tensor defines the torsion scalar, which plays the role of the gravitational Lagrangian,
\begin{equation}
T \equiv -\,S_{\rho}{}^{\mu\nu}T^{\rho}{}_{\mu\nu}.
\label{eq:Tdef}
\end{equation}
The overall sign of $T$ differs between authors. As will become clear, it matters for the physical meaning of the coupling constants. We fix it by requiring that the torsion scalar and the Ricci scalar $\mathring{R}$ of the Levi--Civita connection be related by
\begin{equation}
\mathring{R} = T - B, \qquad B \equiv \frac{2}{e}\,\partial_\mu\!\left(e\,T^{\nu\mu}{}_{\nu}\right),
\label{eq:RTB}
\end{equation}
where quantities built from the Levi--Civita connection are denoted by an overcircle. Since $B$ is a total divergence, the action $\frac{1}{16\pi}\int d^4x\,e\,T$ differs from the Einstein--Hilbert action only by a boundary term. This is the content of TEGR. With this choice a flat Friedmann--Robertson--Walker universe has $T=-6H^2$, in agreement with the convention of teleparallel cosmology \citep{Harko2014}.

The $f(T,\mathcal{T})$ theories generalize TEGR by promoting the Lagrangian to a function of the torsion scalar and of the trace $\mathcal{T}\equiv\Theta^{\mu}{}_{\mu}$ of the matter energy--momentum tensor \citep{Harko2014},
\begin{equation}
\mathcal{S} = \frac{1}{16\pi}\int d^4x\,e\,f(T,\mathcal{T}) + \int d^4x\,e\,\mathcal{L}_m,
\qquad
\Theta_{\mu\nu} = -\frac{2}{\sqrt{-g}}\frac{\delta(\sqrt{-g}\,\mathcal{L}_m)}{\delta g^{\mu\nu}} .
\label{eq:action}
\end{equation}
We write $\Theta_{\mu\nu}$ for the energy--momentum tensor to avoid confusion with torsion. Because the action depends on $\mathcal{T}$, the field equations involve the variation of the trace with respect to the metric, and hence the choice of matter Lagrangian. For stellar matter we take a perfect fluid with energy density $\varepsilon$ and pressure $p$,
\begin{equation}
\Theta_{\mu\nu} = (\varepsilon+p)\,u_\mu u_\nu + p\,g_{\mu\nu},\qquad
\Theta^{\mu}{}_{\nu} = \mathrm{diag}(-\varepsilon,p,p,p),\qquad
\mathcal{T} = -\varepsilon+3p,
\label{eq:fluid}
\end{equation}
and the standard choice $\mathcal{L}_m = p$. The general identity of \citep{HarkoFRT2011}, together with $g^{\alpha\beta}\delta\Theta_{\alpha\beta}/\delta g^{\mu\nu}=-2\Theta_{\mu\nu}+p\,g_{\mu\nu}$ for this Lagrangian, then gives
\begin{equation}
\frac{\delta\mathcal{T}}{\delta g^{\mu\nu}} = \Theta_{\mu\nu}+g^{\alpha\beta}\frac{\delta\Theta_{\alpha\beta}}{\delta g^{\mu\nu}}
= -\Theta_{\mu\nu}+p\,g_{\mu\nu} = -(\varepsilon+p)\,u_\mu u_\nu .
\label{eq:dTrace}
\end{equation}
Varying the action with respect to the tetrad then yields the field equations
\begin{equation}
e^{-1}\partial_\mu\!\left(e\,e_{a}{}^{\rho}S_{\rho}{}^{\mu\nu}f_T\right)
- f_T\,e_{a}{}^{\lambda}T^{\rho}{}_{\mu\lambda}S_{\rho}{}^{\nu\mu}
+ \frac{1}{4}e_{a}{}^{\nu}f
+ \frac{f_{\mathcal{T}}}{2}\left(e_{a}{}^{\lambda}\Theta_{\lambda}{}^{\nu}-p\,e_{a}{}^{\nu}\right)
= -4\pi\,e_{a}{}^{\lambda}\Theta_{\lambda}{}^{\nu},
\label{eq:fieldeqs}
\end{equation}
with $f_T=\partial f/\partial T$ and $f_{\mathcal{T}}=\partial f/\partial\mathcal{T}$. For $f=T$ the left-hand side reduces identically to $-\tfrac12 e_a{}^{\lambda}\mathring{G}_{\lambda}{}^{\nu}$, and Einstein's equations are recovered.

Two features of Eq.~\eqref{eq:fieldeqs} deserve emphasis. First, in the signature adopted here the matter coupling enters through the combination $\Theta^{\nu}{}_{\lambda}-p\,\delta^{\nu}_{\lambda}$. By Eq.~\eqref{eq:dTrace} it is proportional to $(\varepsilon+p)$. The combination $\Theta^{\nu}{}_{\lambda}+p\,\delta^{\nu}_{\lambda}$ of the original formulation \citep{Harko2014} corresponds to the opposite signature with $\mathcal{L}_m=-p$. Carrying it over to the signature $(-,+,+,+)$, as done in earlier studies of compact stars \citep{PaceSaid2017a,PaceSaid2017b,Tangphati2026}, produces effective sources proportional to $(3\varepsilon-5p)$ and $(\varepsilon-7p)$. These are not consistent with $\mathcal{L}_m=p$. Second, the antisymmetric part of Eq.~\eqref{eq:fieldeqs} is proportional to $\partial_\mu f_T$. Whenever $f_{TT}\neq0$ it constrains the tetrad, which therefore cannot be chosen freely \citep{TamaniniBoehmer2012,KrssakSaridakis2016}.

In this work we consider the model
\begin{equation}
f(T,\mathcal{T}) = T + \alpha T^2 + \beta\,\mathcal{T},
\label{eq:model}
\end{equation}
where $\alpha$ has dimensions of length squared and $\beta$ is dimensionless. GR is recovered for $\alpha=\beta=0$. The two corrections enter the theory in very different ways. The trace coupling contributes only through $f_{\mathcal{T}}=\beta$ and through the term $\tfrac14 e_a{}^\nu\beta\mathcal{T}$. It can therefore be absorbed into an effective energy--momentum tensor,
\begin{equation}
8\pi\,\Theta^{\rm eff}_{\mu\nu} = 8\pi\Theta_{\mu\nu} + \beta\left(\Theta_{\mu\nu}-p\,g_{\mu\nu}\right) + \frac{\beta}{2}\,\mathcal{T}g_{\mu\nu},
\label{eq:Theff}
\end{equation}
which for the fluid~\eqref{eq:fluid} is again a perfect fluid with
\begin{equation}
\varepsilon_{\rm eff} = \varepsilon + \frac{\beta}{16\pi}(3\varepsilon-p),\qquad
p_{\rm eff} = p - \frac{\beta}{16\pi}(\varepsilon-3p),\qquad
\varepsilon_{\rm eff}+p_{\rm eff} = \Big(1+\frac{\beta}{8\pi}\Big)(\varepsilon+p).
\label{eq:effective}
\end{equation}
Since $T$ and $\mathring{R}$ differ only by a boundary term, and the trace coupling involves the tetrad only through the metric, the purely linear model ($\alpha=0$) is dynamically equivalent to $f(\mathring{R},\mathcal{T})=\mathring{R}+2\lambda\mathcal{T}$ gravity \citep{HarkoFRT2011} with $\lambda=\beta/2$. Its white-dwarf configurations were studied in \citep{Carvalho2017}. The purely quadratic model ($\beta=0$) is the $f(T)$ model studied in \citet{IlijicSossich2018,Otoniel2026}. The quadratic term is what makes the model genuinely teleparallel. Because $f_{TT}=2\alpha\neq0$, the choice of tetrad becomes physical, and the theory departs from its curvature-based counterparts.

A direct consequence of the trace coupling is that ordinary matter is not conserved. For the static tetrad used below the antisymmetric equations are satisfied identically. The gravitational side of Eq.~\eqref{eq:fieldeqs} is then divergence-free, and the effective tensor~\eqref{eq:Theff} is conserved. Written in terms of the physical fluid, this reads
\begin{equation}
\mathring{\nabla}^{\mu}\Theta_{\mu\nu} = \frac{\beta}{2(8\pi+\beta)}\,\partial_\nu(\varepsilon-p).
\label{eq:nonconservation}
\end{equation}
As we show next, this modifies the condition of hydrostatic equilibrium in a way that is particularly important for white dwarfs.

\section{Stellar structure equations}
\label{sec:structure}

We describe the interior of a white dwarf by the static, spherically symmetric line element
\begin{equation}
ds^2 = -e^{A(r)}dt^2 + e^{B(r)}dr^2 + r^2\left(d\theta^2+\sin^2\theta\,d\phi^2\right).
\label{eq:lineelement}
\end{equation}
The diagonal tetrad associated with this metric is not suitable in the pure-tetrad formulation. It forces $f_{TT}T'=0$ through the antisymmetric field equations and would thus exclude the quadratic term. We therefore adopt the rotated tetrad \citep{TamaniniBoehmer2012,PaceSaid2017a}
\begin{equation}
e^{a}{}_{\mu} =
\begin{pmatrix}
e^{A/2} & 0 & 0 & 0 \\
0 & e^{B/2}\sin\theta\cos\phi & r\cos\theta\cos\phi & -r\sin\theta\sin\phi \\
0 & e^{B/2}\sin\theta\sin\phi & r\cos\theta\sin\phi & r\sin\theta\cos\phi \\
0 & e^{B/2}\cos\theta & -r\sin\theta & 0
\end{pmatrix},
\label{eq:tetrad}
\end{equation}
for which all off-diagonal equations vanish for an arbitrary function $f$. The only non-vanishing torsion components are $T^{t}{}_{rt}=A'/2$ and $T^{\theta}{}_{r\theta}=T^{\phi}{}_{r\phi}=b(r)$, with
\begin{equation}
b(r)\equiv\frac{1-e^{B/2}}{r},
\label{eq:b}
\end{equation}
and the torsion scalar~\eqref{eq:Tdef} becomes
\begin{equation}
T(r) = 2e^{-B}\,b\,(A'+b) = \frac{2e^{-B}}{r^2}\left(1-e^{B/2}\right)\left(1-e^{B/2}+rA'\right),
\label{eq:Tr}
\end{equation}
where a prime denotes $d/dr$. Inside a star $b<0$ while $A'+b>0$, so the torsion scalar is negative. The same field equations follow in the covariant formulation from the diagonal tetrad with a non-vanishing inertial spin connection \citep{KrssakSaridakis2016}. Indeed, Eq.~\eqref{eq:Tr} coincides with the torsion scalar used in \citep{IlijicSossich2018,Otoniel2026}. Our $\alpha$ equals the coupling $\Xi$ of \citep{Otoniel2026} and half the coupling of \citet{IlijicSossich2018,DeBenedictisIlijic2016}.

Introducing the mass function through $e^{-B}=1-2m/r$ and using $f_T=1+2\alpha T$, the $tt$, $rr$ and $\theta\theta$ components of Eq.~\eqref{eq:fieldeqs} take the form
\begin{align}
(1+2\alpha T)\,\frac{2m'}{r^2} - \frac{\alpha}{2}T^2 - 4\alpha\,e^{-B}\,b\,T' &= 8\pi\varepsilon_{\rm eff}, \label{eq:tt}\\
(1+2\alpha T)\left(\frac{e^{-B}A'}{r}-\frac{2m}{r^3}\right) + \frac{\alpha}{2}T^2 &= 8\pi p_{\rm eff}, \label{eq:rr}\\
\frac{e^{-B}}{4}\Big[f_T\big(2A''+A'^2-A'B'\big)+2A'f_T'\Big]
+\frac{e^{-B}}{2r}\Big[f_T(3A'-B')+2f_T'\Big]
-\frac{e^{-B/2}}{r}\big(f_TA'+f_T'\big) & \nonumber\\
+\frac{f_T}{r^2}\big(1-e^{-B/2}\big)^2 - \frac{T+\alpha T^2}{2} &= 8\pi p_{\rm eff}, \label{eq:thth}
\end{align}
with $f_T'=2\alpha T'$. Apart from the explicit $\alpha T^2$ terms, the quadratic term enters in two ways. The factor $f_T$ acts as a position-dependent rescaling of the gravitational coupling. The derivative $T'$ in the $tt$ equation has no counterpart in GR. For $\alpha=0$ these equations reduce to Einstein's equations sourced by the effective fluid~\eqref{eq:effective}.

The hydrostatic equilibrium condition follows from the radial component of Eq.~\eqref{eq:nonconservation}, $p_{\rm eff}'+\tfrac12(\varepsilon_{\rm eff}+p_{\rm eff})A'=0$. For a barotropic equation of state with squared sound speed $c_s^2\equiv dp/d\varepsilon$, it gives
\begin{equation}
\frac{dp}{dr} = -\frac{(8\pi+\beta)(\varepsilon+p)}{16\pi+3\beta-\beta/c_s^2}\,A' ,
\label{eq:dpdr}
\end{equation}
which reduces to the familiar relation $p'=-\tfrac12(\varepsilon+p)A'$ for $\beta=0$. Equations~\eqref{eq:tt}, \eqref{eq:rr} and \eqref{eq:dpdr} constitute the modified Tolman--Oppenheimer--Volkoff system. Since $T$ is linear in $A'$, the $rr$ equation is a quadratic for $A'$. We take the root that reduces to the GR value $A'=(2m+8\pi r^3p_{\rm eff})/[r(r-2m)]$ as $\alpha\to0$. Once $A''$ is eliminated using the derivative of the $rr$ equation, the $tt$ equation is linear in $m'$. The system therefore closes in the variables $p(r)$ and $m(r)$, while the $\theta\theta$ equation is satisfied automatically. For $\alpha>0$ the coefficient of $m'$ can vanish at high density. Numerically, this coincides with the condition $f_T+2Tf_{TT}=1+6\alpha T=0$. The same combination appears in the denominator of the modified Friedmann equations of $f(T)$ cosmology \citep{CaiReview2016,Nunes:2016qyp}. In all our sequences it occurs only on the unstable branch, beyond the maximum mass.

The appearance of the sound speed in the denominator of Eq.~\eqref{eq:dpdr} is the central new feature of the trace coupling. The denominator must remain positive throughout the star. Otherwise the pressure gradient either diverges or acquires the wrong sign. In degenerate matter $c_s^2$ is small even in the core and decreases outwards. For $\beta>0$ the denominator of Eq.~\eqref{eq:dpdr} therefore vanishes where $c_s^2=\beta/(16\pi+3\beta)$. Somewhat deeper, the effective pressure already reaches zero, at $p\simeq\beta\varepsilon/16\pi$, while the electron pressure is still finite. For the cold equation of state these two points lie at $\rho\approx2\times10^{2}$ and $4\times10^{2}\,\mathrm{g\,cm^{-3}}$ for $\beta=10^{-5}$, and at $6\times10^{3}$ and $1.4\times10^{4}\,\mathrm{g\,cm^{-3}}$ for $\beta=10^{-4}$. They can therefore lie outside the surface density adopted below. A real white dwarf, however, continues through a non-degenerate envelope up to the photosphere, where the sound speed is thermal, $c_s^2\sim kT/(\mu m_uc^2)\sim10^{-9}$ for $T\sim10^{4}\,\mathrm{K}$. The effective pressure then vanishes below the photosphere while the gas pressure is still finite, so the star cannot support its observed atmosphere unless $\beta\lesssim10^{-7}$, a value without any effect on the structure. Observed white dwarfs therefore require
\begin{equation}
\beta\le 0 ,
\label{eq:betarange}
\end{equation}
in line with the restriction found in $f(R,\mathcal{T})$ gravity for a surface at $p=0$ \citep{Carvalho2017}. For $\beta<0$ the term $|\beta|/c_s^2$ dominates near the surface. Equation~\eqref{eq:dpdr} then reduces to $d\varepsilon/dr\simeq-(8\pi+\beta)\,\varepsilon A'/|\beta|$. The density falls off roughly exponentially, with scale height $\ell\simeq|\beta|/(8\pi A')$, instead of vanishing at a finite radius. The mass converges rapidly in these outer layers, but a radius defined by $p=0$ is not well defined (Appendix~\ref{app:surface}). The cold degenerate equation of state is not valid in the outer, non-degenerate envelope anyway. We therefore define the radius $R$ as the point where the rest-mass density drops to $\rho_s=10^4\,\mathrm{g\,cm^{-3}}$, and we examine the dependence on $\rho_s$ in Sec.~\ref{sec:obs}.

The quadratic term has a second consequence. Outside the star $\varepsilon=p=0$ and the trace coupling disappears. The torsion scalar, however, does not vanish, so the $\alpha T^2$ terms persist. The Schwarzschild metric is therefore not a vacuum solution \citep{DeBenedictisIlijic2016,IlijicSossich2018}. The mass function keeps evolving beyond $R$, and the gravitational mass is the asymptotic value
\begin{equation}
M \equiv M_\infty = \lim_{r\to\infty}m(r),
\label{eq:Minf}
\end{equation}
as advocated in \citep{deAraujoFortes2023} and adopted in \citep{Otoniel2026}. We obtain it by continuing the integration into the vacuum region. The size of the effect follows from the weak-field vacuum solution. Outside the star $b\simeq-m/r^2$ and $A'\simeq2m/r^2$, so Eq.~\eqref{eq:Tr} gives $T\simeq-2m^2/r^4$. To first order in $\alpha$ the vacuum $tt$ equation becomes $m'\simeq2\alpha r^2bT'\simeq-16\alpha m^3/r^5$. Integrating from the surface to infinity yields
\begin{equation}
M_\infty - m(R) \simeq -\frac{4\alpha M^3}{R^4},
\qquad\text{or equivalently}\qquad
\frac{M_\infty-m(R)}{m(R)} \simeq 2\alpha T(R).
\label{eq:Mshift}
\end{equation}
This is the $1/r^5$ tail of $g_{rr}$ contained in the perturbative vacuum solution of \citep{DeBenedictisIlijic2016}. Since $T<0$, a positive $\alpha$ lowers the mass seen by a distant observer. The integration constant in $A(r)$ is fixed by $e^{A}\to1$ at infinity, so that the surface redshift is $z=e^{-A(R)/2}-1$.
For the numerical solutions we restore physical units through $\varepsilon\to G\varepsilon/c^4$, $p\to Gp/c^4$ and $m\to Gm/c^2$, keeping $\alpha$ in $\mathrm{cm^2}$ and $\beta$ dimensionless.
\subsection{Equation of state}
\label{sec:eos}

White dwarfs are supported against gravity by the pressure of a cold, fully degenerate electron gas. We describe it by the Chandrasekhar equation of state \citep{Chandrasekhar1935,ShapiroTeukolsky1983}. In terms of the relativity parameter $x=p_F/(m_ec)$, where $p_F$ is the electron Fermi momentum, the electron number density and the rest-mass density are
\begin{equation}
n_e = \frac{x^3}{3\pi^2\lambda_e^3},\qquad \rho=\mu_e m_u n_e,
\label{eq:ne_rho}
\end{equation}
with $\lambda_e=\hbar/(m_ec)$, $m_u$ the atomic mass unit and $\mu_e=2$, appropriate for helium, carbon and oxygen. The pressure and energy density of the electron gas are
\begin{align}
p &= \frac{\pi m_e^4c^5}{3h^3}\left[x(2x^2-3)\sqrt{x^2+1}+3\sinh^{-1}x\right], \label{eq:pe}\\
\varepsilon_e &= \frac{\pi m_e^4c^5}{h^3}\left[x(1+2x^2)\sqrt{x^2+1}-\sinh^{-1}x\right], \label{eq:epse}
\end{align}
and the total energy density, $\varepsilon=\varepsilon_e+\rho c^2$, is dominated by the rest mass of the ions. The squared sound speed that enters Eq.~\eqref{eq:dpdr} follows from
\begin{equation}
c_s^2 = \frac{dp/dx}{d\varepsilon/dx},\qquad
\frac{dp}{dx}=\frac{8\pi m_e^4c^5}{3h^3}\frac{x^4}{\sqrt{1+x^2}},\qquad
\frac{d\varepsilon}{dx}=\frac{8\pi m_e^4c^5}{h^3}x^2\sqrt{1+x^2}+\frac{3\rho c^2}{x}.
\label{eq:cs2}
\end{equation}
Because $\varepsilon\simeq\rho c^2$, one has $c_s^2\simeq\frac{1}{3}\frac{m_e}{\mu_em_u}\frac{x^2}{\sqrt{1+x^2}}$. It does not exceed $\sim2\times10^{-3}$ in white dwarfs and becomes very small near the surface. This smallness of the sound speed, a characteristic feature of degenerate electron matter, makes white dwarfs particularly sensitive to the trace coupling. 

The equation of state above ignores electron captures on nuclei, which set an upper density for a given composition. Capture on a nucleus $(A,Z)$ starts when the electron chemical potential, including the rest mass, reaches $Q+m_ec^2$. Here $Q$ is the atomic mass difference between the daughter and parent nuclei. This gives thresholds of $13.88\,\mathrm{MeV}$ for $^{12}$C and $10.93\,\mathrm{MeV}$ for $^{16}$O. For the equation of state above, the corresponding densities are $\rho_\beta=3.90\times10^{10}\,\mathrm{g\,cm^{-3}}$ for carbon and $1.90\times10^{10}\,\mathrm{g\,cm^{-3}}$ for oxygen, in agreement with \citep{ShapiroTeukolsky1983}. Sequences without an earlier turning point are stopped at $\rho_\beta(^{12}\mathrm{C})$. For validation we also use a variant with the Coulomb lattice correction of a body-centred cubic $^{12}$C crystal \citep{Salpeter1961}, described in Appendix~\ref{app:coulomb}.

\section{Results and discussion}
\label{sec:results}

\subsection{Numerical method and validation}
\label{sec:validation}

For given values of $(\alpha,\beta)$ and of the central density $\rho_c$, we integrate the structure equations outward from the centre, with $m(0)=0$ and $p(0)=p_c$, up to the surface $\rho=\rho_s$. We then continue the integration in vacuum to $1000\,R$ to obtain $M_\infty$. We use an eighth-order Runge--Kutta (DOP853) integrator with a relative tolerance of $10^{-10}$. The $\theta\theta$ equation~\eqref{eq:thth}, which is not used in the integration, is satisfied to better than $3\times10^{-9}$ relative to $8\pi\varepsilon_{\rm eff}$. Repeating the integration for $10^5\le\rho_c\le10^{11}\,\mathrm{g\,cm^{-3}}$ (or up to $\rho_\beta$, see Sec.~\ref{sec:eos}) yields sequences of equilibrium configurations, along which we locate the maximum-mass star.

The two single-coupling limits of our model have been studied before, and we use them as tests (Table~\ref{tab:validation}). In the GR limit we obtain $M_{\max}=1.4158\,M_\odot$ at $\rho_c=2.311\times10^{10}\,\mathrm{g\,cm^{-3}}$, the general-relativistic Chandrasekhar limit for $\mu_e=2$. \citep{Carvalho2017} reports $1.416\,M_\odot$ at $2.307\times10^{10}\,\mathrm{g\,cm^{-3}}$. For $\alpha=0$ and $\beta=-8\times10^{-4}$ we find $M_{\max}=1.4671\,M_\odot$, the $f(R,\mathcal{T})$ result of \citep{Carvalho2017} for $\lambda=\beta/2$. For $\beta=0$ we compare with the quadratic $f(T)$ results of \citep{Otoniel2026}, using their Coulomb-corrected equation of state. We reproduce their GR maximum mass to $0.05\%$ and their first maximum for $\alpha=10^{18}\,\mathrm{cm^2}$ to $0.03\%$. Their relative changes in mass and radius at $\rho_\beta$ agree with ours to $0.02$ percentage points (Appendix~\ref{app:coulomb}). The radii of \citep{Carvalho2017} differ from ours because of the surface definition, as explained in Appendix~\ref{app:surface}. Unless stated otherwise, all results below use the ideal equation of state of Sec.~\ref{sec:eos}.
\begin{table*}[htbp]
\centering
\caption{Validation against the single-coupling limits of the model. Ours: $M_\infty$, with $R$ at $\rho_s=10^4\,\mathrm{g\,cm^{-3}}$. \citep{Carvalho2017} uses $f=R+2\lambda\mathcal{T}$ ($\beta=2\lambda$) and a surface at $p=0$. \citep{Otoniel2026} uses a Coulomb-corrected $^{12}$C equation of state with the electron-capture cut; for these rows we use the same equation of state (Appendix~\ref{app:coulomb}). Masses in $M_\odot$, densities in $\mathrm{g\,cm^{-3}}$, radii in km.}
\label{tab:validation}
{\small\setlength{\tabcolsep}{5pt}
\begin{tabular}{llccc}
\toprule
case & quantity & ours & reference & deviation \\
\midrule
GR, ideal & $M_{\max}$ & 1.4158 & 1.416 \citep{Carvalho2017} & $-0.01\%$ \\
 & $\rho_c$ & $2.311\times10^{10}$ & $2.307\times10^{10}$ & $+0.2\%$ \\
 & $R$ & 1026.8 & 1021 & surface$^a$ \\
$\beta=-8\times10^{-4}$, ideal & $M_{\max}$ & 1.4671 & 1.467 \citep{Carvalho2017} & $+0.01\%$ \\
 & $\rho_c$ & $3.091\times10^{9}$ & $3.366\times10^{9}$ & $-8.2\%^b$ \\
 & $R$ & 2075 & 2970 & surface$^a$ \\
GR, Coulomb & $M_{\max}$ & 1.3843 & 1.385 \citep{Otoniel2026} & $-0.05\%$ \\
 & $M$, $R$ at $\rho_c=2.708\times10^{10}$ & 1.3842, 972.4$^c$ & 1.385, 972.7 & $-0.06\%$, $-0.03\%$ \\
$\alpha=10^{18}\,\mathrm{cm^2}$, Coulomb & first max.: $M$, $\rho_c$ & 1.36566, $6.010\times10^{9}$ & 1.366, $6.01\times10^{9}$ & $-0.025\%$, $<0.01\%$ \\
$\alpha=-10^{18}\,\mathrm{cm^2}$, Coulomb & $M(\rho_\beta)$ & 1.5069 & 1.508 & $-0.07\%$ \\
$\rho_c=\rho_\beta$, Coulomb & $\Delta M/M_{\rm GR}$ ($\alpha=\mp10^{18}$) & $+8.95\%$, $-10.11\%$ & $+8.95\%$, $-10.12\%$ & $\le0.01$\,pp \\
 & $\Delta R/R_{\rm GR}$ ($\alpha=\mp10^{18}$) & $-1.04\%$, $+2.89\%$ & $-1.05\%$, $+2.91\%$ & $\le0.02$\,pp \\
\bottomrule
\end{tabular}}
\par\smallskip{\footnotesize $^a$See Appendix~\ref{app:surface}. $^b$The maximum is very flat in $\rho_c$. $^c$$R$ at $p/p_c=10^{-12}$. The Coulomb GR maximum lies at $\rho_c=2.39\times10^{10}$; between this density and $2.708\times10^{10}$ the mass changes by only $6\times10^{-5}\,M_\odot$.}
\end{table*}
\subsection{Internal and external structure}

\begin{figure*}[htbp]
\centering
\includegraphics[width=\textwidth]{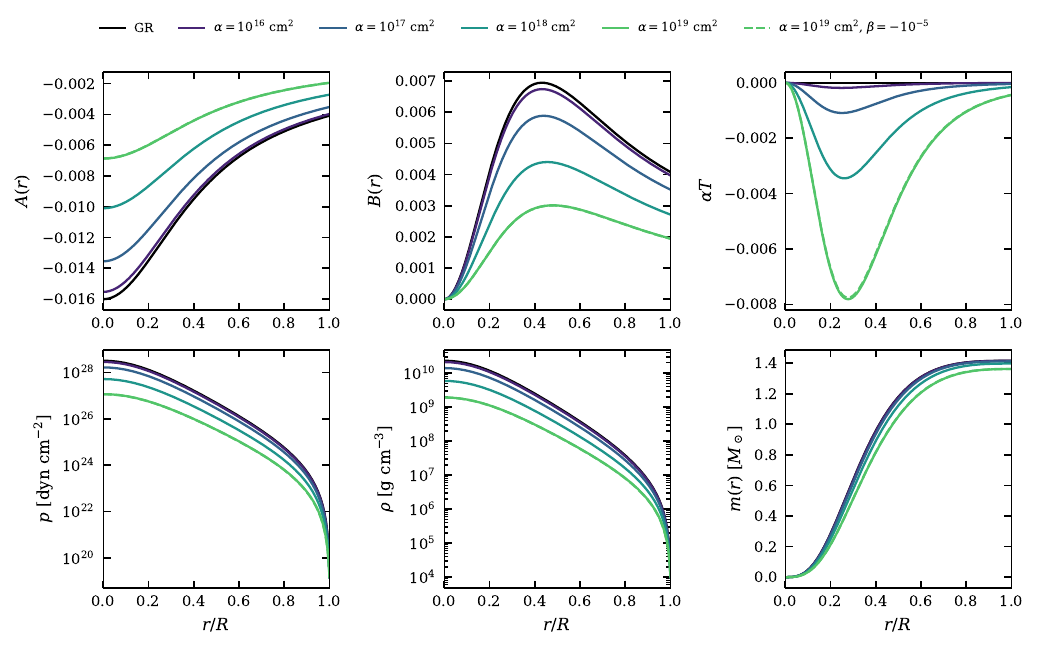}
\caption{Radial profiles of the maximum-mass white dwarfs for $\alpha=0$ (GR), $10^{16}$, $10^{17}$, $10^{18}$ and $10^{19}\,\mathrm{cm^2}$ with $\beta=0$. Panels: the metric functions $A(r)$, normalized so that $e^A\to1$ at infinity, and $B(r)$; the correction $\alpha T$; the pressure $p$; the rest-mass density $\rho$; and the mass function $m(r)$. Dashed: the combined star $(\alpha,\beta)=(10^{19}\,\mathrm{cm^2},-10^{-5})$, with $M_{\max}=1.3611\,M_\odot$.}
\label{fig:struct_alpha}
\end{figure*}

Figure~\ref{fig:struct_alpha} shows the radial profiles of the maximum-mass configurations for $\alpha$ between $0$ and $10^{19}\,\mathrm{cm^2}$, with $\beta=0$. As $\alpha$ increases, the central density and pressure of the maximum-mass star drop by an order of magnitude. The star becomes correspondingly more extended. The metric potentials follow this trend: $A(r)$ and $B(r)$ become shallower, with $|A(0)|$ decreasing from $1.6\times10^{-2}$ in GR to $6.9\times10^{-3}$ for $\alpha=10^{19}\,\mathrm{cm^2}$. The dimensionless correction $\alpha T$ is negative everywhere. It vanishes at the centre, decays towards the surface, and reaches its largest magnitude, $\alpha T\simeq-7.8\times10^{-3}$ for $\alpha=10^{19}\,\mathrm{cm^2}$, at $r/R\simeq0.27$. Since $f_T=1+2\alpha T$, the quadratic term weakens the effective gravitational coupling by at most $\sim1.6\%$. It does so mainly in the intermediate layers where most of the mass resides. Although modest, this weakening is enough to change the location and value of the maximum mass. The dashed line shows a combined star with $(\alpha,\beta)=(10^{19}\,\mathrm{cm^2},-10^{-5})$. It nearly coincides with the $\alpha$-only star, because a trace coupling of this size hardly affects the maximum-mass configuration.

\begin{figure*}[htbp]
\centering
\includegraphics[width=\textwidth]{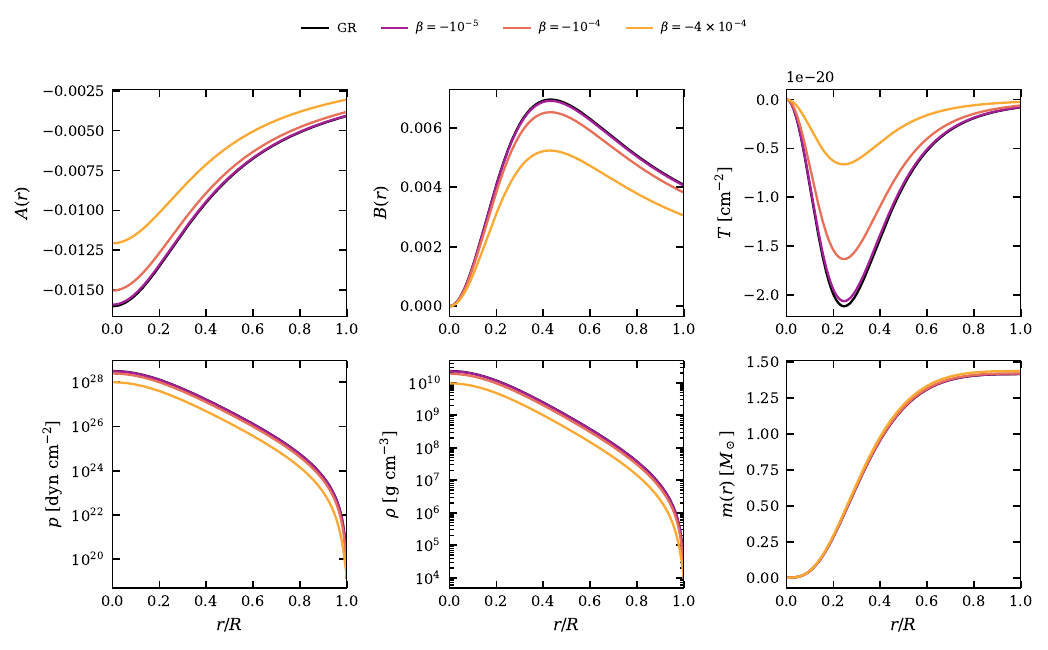}
\caption{As Fig.~\ref{fig:struct_alpha} for $\beta=0$, $-10^{-5}$, $-10^{-4}$ and $-4\times10^{-4}$ with $\alpha=0$. The third panel shows the torsion scalar $T(r)$ of Eq.~\eqref{eq:Tr}.}
\label{fig:struct_beta}
\end{figure*}

The trace coupling acts through a different mechanism, although its effect on the maximum-mass star is qualitatively similar (Fig.~\ref{fig:struct_beta}). By Eq.~\eqref{eq:effective}, a negative $\beta$ reduces the effective inertial density $\varepsilon_{\rm eff}+p_{\rm eff}$ that opposes the pressure gradient. The term $|\beta|/c_s^2$ in Eq.~\eqref{eq:dpdr} further flattens the gradient wherever the sound speed is small. The resulting stars have lower central densities and shallower potentials. The torsion scalar, shown in the third panel, decreases in magnitude accordingly.

\begin{figure*}[htbp]
\centering
\includegraphics[width=\textwidth]{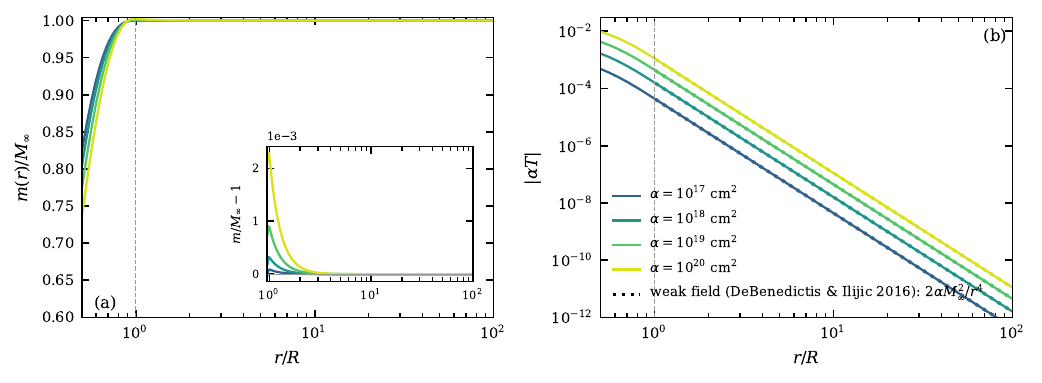}
\caption{Exterior of the maximum-mass stars for $\alpha=10^{17}$--$10^{20}\,\mathrm{cm^2}$ ($\beta=0$). (a) Mass function $m(r)/M_\infty$; the inset shows $m/M_\infty-1$ outside the star. (b) $|\alpha T|$ together with the weak-field result $2\alpha M_\infty^2/r^4$ (dotted), which follows from the vacuum solution of \citep{DeBenedictisIlijic2016}.}
\label{fig:exterior}
\end{figure*}

The quadratic term also leaves an imprint outside the star. Figure~\ref{fig:exterior} shows that $m(r)$ keeps decreasing beyond the surface over a few stellar radii before settling to $M_\infty$. Meanwhile $|\alpha T|$ follows the weak-field law $2\alpha M_\infty^2/r^4$ closely already at $r\simeq2R$. The numerical mass tail agrees with $4\alpha M^3/r^4$ to $2\times10^{-4}$ at $r=10R$. The resulting mass correction is listed in Table~\ref{tab:ext}. It grows from $9\times10^{-5}$ for $\alpha=10^{17}\,\mathrm{cm^2}$ to $2.2\times10^{-3}$ for $\alpha=10^{20}\,\mathrm{cm^2}$, and agrees with Eq.~\eqref{eq:Mshift} to better than $0.6\%$. The correction exceeds $10^{-4}\,M_\odot$ for $\alpha\gtrsim10^{17}\,\mathrm{cm^2}$. This is comparable to the shift of the maximum mass itself. Identifying $m(R)$ with the gravitational mass would therefore bias the results, as noted for neutron stars in \citep{deAraujoFortes2023}. Equation~\eqref{eq:Mshift} gives a simple estimate of this bias for any star with $f_{TT}\neq0$.
\begin{table}[htbp]
\centering
\caption{Exterior mass correction for the maximum-mass stars ($\beta=0$). Numerical $(M_\infty-m(R))/m(R)$, compared with $2\alpha T(R)$ and with the weak-field estimate $-4\alpha M^2/R^4$ of Eq.~\eqref{eq:Mshift}. The last column gives the ratio of the numerical value to the weak-field estimate.}
\label{tab:ext}
\begin{tabular}{lccccc}
\toprule
$\alpha$ [cm$^2$] & $M_\infty$ [$M_\odot$] & $(M_\infty-m(R))/m(R)$ & $2\alpha T(R)$ & $-4\alpha M^2/R^4$ & ratio \\
\midrule
$10^{17}$ & 1.4115 & $-8.747\times10^{-5}$ & $-8.754\times10^{-5}$ & $-8.722\times10^{-5}$ & 1.0029 \\
$10^{18}$ & 1.3970 & $-3.176\times10^{-4}$ & $-3.178\times10^{-4}$ & $-3.167\times10^{-4}$ & 1.0027 \\
$10^{19}$ & 1.3601 & $-8.839\times10^{-4}$ & $-8.843\times10^{-4}$ & $-8.810\times10^{-4}$ & 1.0033 \\
$10^{20}$ & 1.2846 & $-2.212\times10^{-3}$ & $-2.213\times10^{-3}$ & $-2.200\times10^{-3}$ & 1.0055 \\
\bottomrule
\end{tabular}
\end{table}
\subsection{Mass--radius relation and maximum mass}
\label{sec:MR}

\begin{figure*}[htbp]
\centering
\includegraphics[width=\textwidth]{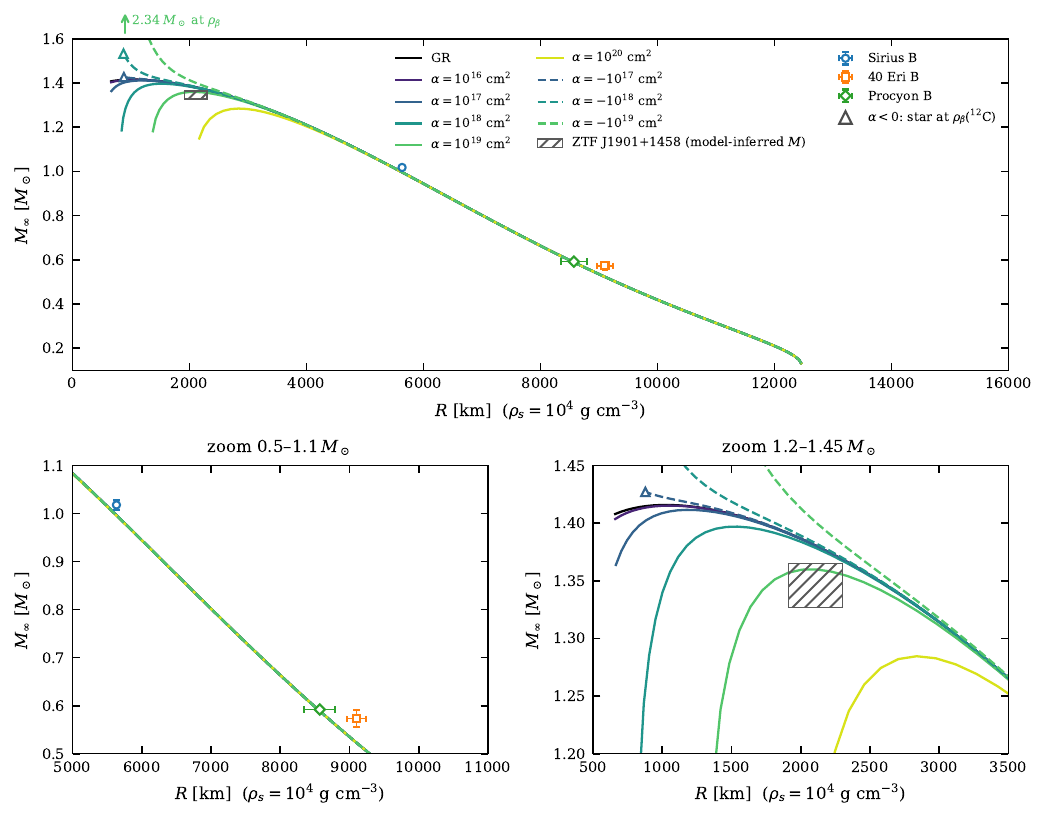}
\caption{Mass--radius relations for $f=T+\alpha T^2$ ($\beta=0$). Solid: $\alpha=0$ (GR), $10^{16}$, $10^{17}$, $10^{18}$, $10^{19}$ and $10^{20}\,\mathrm{cm^2}$. Dashed: $\alpha=-10^{17}$, $-10^{18}$ and $-10^{19}\,\mathrm{cm^2}$, which have no turning point and end at $\rho_\beta(^{12}\mathrm{C})$ (open triangles; the $\alpha=-10^{19}\,\mathrm{cm^2}$ sequence ends at $2.34\,M_\odot$, above the panel). The lower panels zoom into $0.5$--$1.1\,M_\odot$ and $1.2$--$1.45\,M_\odot$. Points with $1\sigma$ error bars: Sirius~B, 40~Eri~B and Procyon~B, whose masses are dynamical. Hatched box: ZTF~J1901+1458, whose mass is model-inferred.}
\label{fig:MR_alpha}
\end{figure*}

The global effect of the quadratic torsion term is shown in Fig.~\ref{fig:MR_alpha} and summarized in Table~\ref{tab:mmax_alpha}. For $\alpha>0$ the maximum mass decreases monotonically. It falls from $1.4158\,M_\odot$ in GR to $1.4115$, $1.3970$, $1.3601$ and $1.2846\,M_\odot$ for $\alpha=10^{17}$, $10^{18}$, $10^{19}$ and $10^{20}\,\mathrm{cm^2}$. The corresponding central density falls from $2.3\times10^{10}$ to $5.6\times10^{8}\,\mathrm{g\,cm^{-3}}$, and the radius grows from about $1030$ to $2840\,\mathrm{km}$. The surface redshift decreases accordingly, reflecting the lower compactness of these stars. For $\alpha\lesssim10^{16}\,\mathrm{cm^2}$ the maximum mass changes by less than $10^{-3}\,M_\odot$.
\begin{table}[htbp]
\centering
\caption{Maximum-mass configurations for $f=T+\alpha T^2$ ($\beta=0$): central density $\rho_c$, maximum mass $M_{\max}$ (the asymptotic mass $M_\infty$), radius $R$ and surface redshift $z$. The case $\alpha=0$ corresponds to GR. $^\dagger$No turning point below $\rho_\beta(^{12}\mathrm{C})=3.896\times10^{10}\,\mathrm{g\,cm^{-3}}$; the row gives the star at $\rho_\beta$.}
\label{tab:mmax_alpha}
\begin{tabular}{lcccc}
\toprule
$\alpha$ [cm$^2$] & $\rho_c$ [g\,cm$^{-3}$] & $M_{\max}$ [$M_\odot$] & $R$ [km] & $z$ \\
\midrule
$0$ & $2.311\times10^{10}$ & 1.4158 & 1026.8 & $2.042\times10^{-3}$ \\
$10^{15}$ & $2.288\times10^{10}$ & 1.4157 & 1029.9 & $2.036\times10^{-3}$ \\
$10^{16}$ & $2.112\times10^{10}$ & 1.4152 & 1054.5 & $1.988\times10^{-3}$ \\
$5\times10^{16}$ & $1.676\times10^{10}$ & 1.4133 & 1128.8 & $1.854\times10^{-3}$ \\
$10^{17}$ & $1.406\times10^{10}$ & 1.4115 & 1188.1 & $1.759\times10^{-3}$ \\
$5\times10^{17}$ & $7.950\times10^{9}$ & 1.4032 & 1399.7 & $1.484\times10^{-3}$ \\
$10^{18}$ & $5.902\times10^{9}$ & 1.3970 & 1522.6 & $1.358\times10^{-3}$ \\
$5\times10^{18}$ & $2.740\times10^{9}$ & 1.3744 & 1881.5 & $1.080\times10^{-3}$ \\
$10^{19}$ & $1.924\times10^{9}$ & 1.3601 & 2068.6 & $9.723\times10^{-4}$ \\
$5\times10^{19}$ & $8.162\times10^{8}$ & 1.3125 & 2584.4 & $7.508\times10^{-4}$ \\
$10^{20}$ & $5.567\times10^{8}$ & 1.2846 & 2844.0 & $6.676\times10^{-4}$ \\
\midrule
$-10^{16}$ & $2.612\times10^{10}$ & 1.4165 & 990.2 & $2.119\times10^{-3}$ \\
$-10^{17}$\,$^\dagger$ & $3.896\times10^{10}$ & 1.4271 & 877.7 & $2.410\times10^{-3}$ \\
$-10^{18}$\,$^\dagger$ & $3.896\times10^{10}$ & 1.5331 & 870.3 & $2.611\times10^{-3}$ \\
$-10^{19}$\,$^\dagger$ & $3.896\times10^{10}$ & 2.3416 & 898.5 & $3.871\times10^{-3}$ \\
$-10^{20}$\,$^\dagger$ & $3.896\times10^{10}$ & 6.8112 & 1197.9 & $8.503\times10^{-3}$ \\
\bottomrule
\end{tabular}
\end{table}

A notable feature of Fig.~\ref{fig:MR_alpha} is that the modification is confined to the high-mass end of the sequence. Below about $1.2\,M_\odot$ all curves with $\alpha>0$ are indistinguishable from GR. This follows from the structure of the correction: $|\alpha T|\sim\alpha (GM/c^2)^2/R^4$ grows steeply with compactness and is negligible in diffuse stars. For $\alpha\gtrsim5\times10^{16}\,\mathrm{cm^2}$ the maximum lies below both capture densities of Sec.~\ref{sec:eos}. In this regime the mass limit is set by the gravitational modification rather than by the microphysics. The trend of a lower maximum mass reached at lower density and larger radius agrees with \citep{Otoniel2026} for white dwarfs and with \citep{IlijicSossich2018} for polytropes. A qualitatively similar trend was found for $f(Q)=Q+\alpha Q^2$ \citep{Sahoo2026}. This suggests that it is a generic consequence of a positive quadratic correction to the gravitational Lagrangian.

For $\alpha<0$ the behaviour changes qualitatively. For $\alpha\le-10^{17}\,\mathrm{cm^2}$ the mass rises monotonically up to $\rho_\beta$ (Figs.~\ref{fig:MR_alpha} and \ref{fig:Mrhoc}). The exterior correction~\eqref{eq:Mshift}, now positive, raises it further. For $-10^{17}<\alpha<0$ the curve $M(\rho_c)$ is nearly flat close to $\rho_\beta$, and a shallow maximum may or may not appear below $\rho_\beta$. In this range the limiting mass stays within $0.01\,M_\odot$ of the GR value. The structure equations nevertheless remain regular: $f_T>1$ everywhere, and no singularity occurs. The limiting mass is then set by electron capture. At $\rho_\beta$ we find $1.427$, $1.533$ and $2.342\,M_\odot$ for $\alpha=-10^{17}$, $-10^{18}$ and $-10^{19}\,\mathrm{cm^2}$, and $6.8\,M_\odot$ for $-10^{20}\,\mathrm{cm^2}$ (Table~\ref{tab:mmax_alpha}). This agrees with \citep{Otoniel2026}, where the sequences are also monotonic up to capture for $\Xi\le-10^{17}\,\mathrm{cm^2}$. It is also consistent with the larger particle numbers found for negative coupling in \citep{IlijicSossich2018}. Negative $\alpha$ is therefore not excluded by the structure of the solutions. It allows super-Chandrasekhar white dwarfs, which have been discussed as progenitors of overluminous type Ia supernovae \citep{DasMukhopadhyay2015}. In some matter-coupled theories such masses are reached only at $\rho_c\sim10^{12}\,\mathrm{g\,cm^{-3}}$ \citep{Priyobarta2026}, well above $\rho_\beta$. Here, by contrast, these masses are reached at densities no higher than $\rho_\beta$, where electron capture sets the limit.

\begin{figure*}[htbp]
\centering
\includegraphics[width=\textwidth]{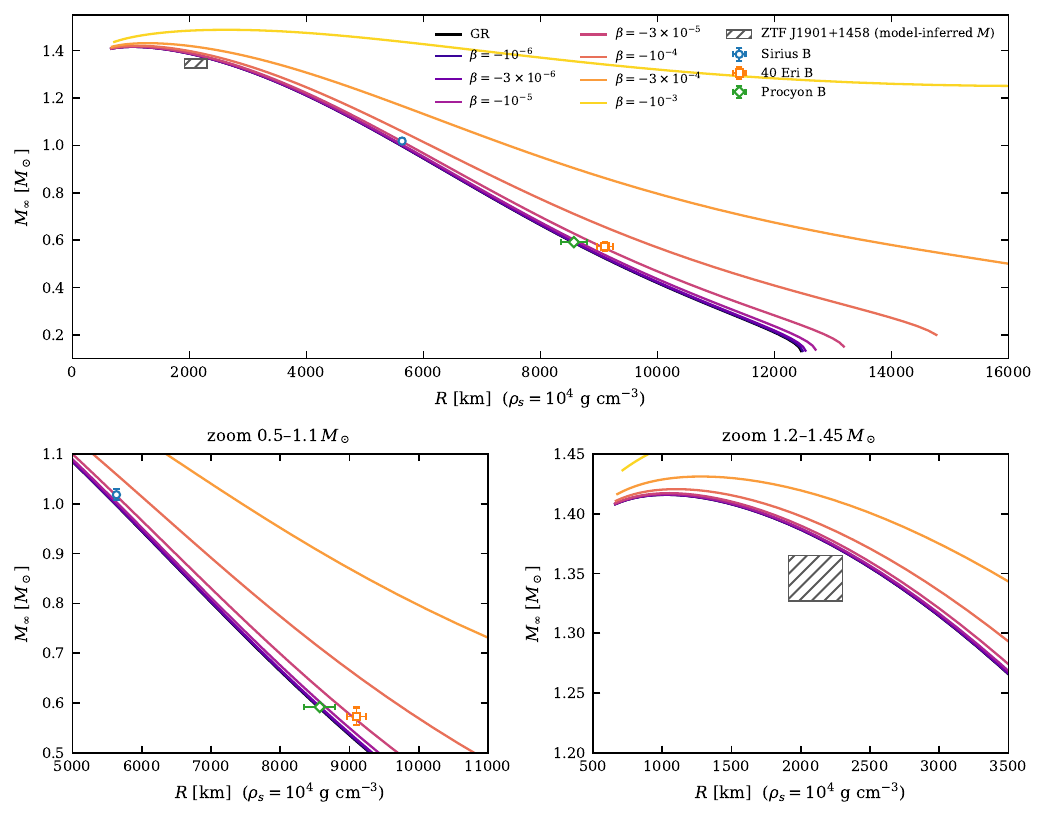}
\caption{As Fig.~\ref{fig:MR_alpha} for $f=T+\beta\mathcal{T}$ ($\alpha=0$) with $\beta=0$, $-10^{-6}$, $-3\times10^{-6}$, $-10^{-5}$, $-3\times10^{-5}$, $-10^{-4}$, $-3\times10^{-4}$ and $-10^{-3}$.}
\label{fig:MR_beta}
\end{figure*}

The trace coupling affects the sequences in the opposite direction, as shown in Fig.~\ref{fig:MR_beta} and Table~\ref{tab:mmax_beta}. A negative $\beta$ raises the maximum mass, to $1.4205$, $1.4371$ and $1.4876\,M_\odot$ for $\beta=-10^{-4}$, $-4\times10^{-4}$ and $-10^{-3}$, again at lower central density and larger radius. More striking is its effect on low-mass white dwarfs. The term $|\beta|/c_s^2$ is largest where the sound speed is smallest. The trace coupling therefore inflates the outer layers of diffuse stars much more than those of compact ones. For $\beta=-10^{-4}$ a $0.6\,M_\odot$ white dwarf is about $14\%$ larger than in GR. For $\beta=-10^{-3}$ no configuration with $\rho_c\ge10^5\,\mathrm{g\,cm^{-3}}$ is lighter than $1.18\,M_\odot$. The two corrections therefore leave distinct signatures. The parameter $\alpha$ acts only on near-Chandrasekhar stars, whereas $\beta$ acts most strongly on ordinary, low-mass white dwarfs. As we show in Sec.~\ref{sec:obs}, this is what allows observations to constrain them separately.
\begin{table}[htbp]
\centering
\caption{As Table~\ref{tab:mmax_alpha} for $f=T+\beta\mathcal{T}$ ($\alpha=0$).}
\label{tab:mmax_beta}
\begin{tabular}{lcccc}
\toprule
$\beta$ & $\rho_c$ [g\,cm$^{-3}$] & $M_{\max}$ [$M_\odot$] & $R$ [km] & $z$ \\
\midrule
$0$ & $2.311\times10^{10}$ & 1.4158 & 1026.8 & $2.042\times10^{-3}$ \\
$-10^{-6}$ & $2.307\times10^{10}$ & 1.4158 & 1027.5 & $2.041\times10^{-3}$ \\
$-3\times10^{-6}$ & $2.298\times10^{10}$ & 1.4159 & 1028.9 & $2.038\times10^{-3}$ \\
$-10^{-5}$ & $2.268\times10^{10}$ & 1.4163 & 1033.7 & $2.029\times10^{-3}$ \\
$-3\times10^{-5}$ & $2.183\times10^{10}$ & 1.4172 & 1047.6 & $2.004\times10^{-3}$ \\
$-10^{-4}$ & $1.902\times10^{10}$ & 1.4205 & 1099.5 & $1.913\times10^{-3}$ \\
$-3\times10^{-4}$ & $1.230\times10^{10}$ & 1.4311 & 1280.5 & $1.655\times10^{-3}$ \\
$-4\times10^{-4}$ & $9.654\times10^{9}$ & 1.4371 & 1393.4 & $1.526\times10^{-3}$ \\
$-10^{-3}$ & $1.579\times10^{9}$ & 1.4876 & 2629.0 & $8.366\times10^{-4}$ \\
\bottomrule
\end{tabular}
\end{table}

\begin{figure*}[htbp]
\centering
\includegraphics[width=0.9\textwidth]{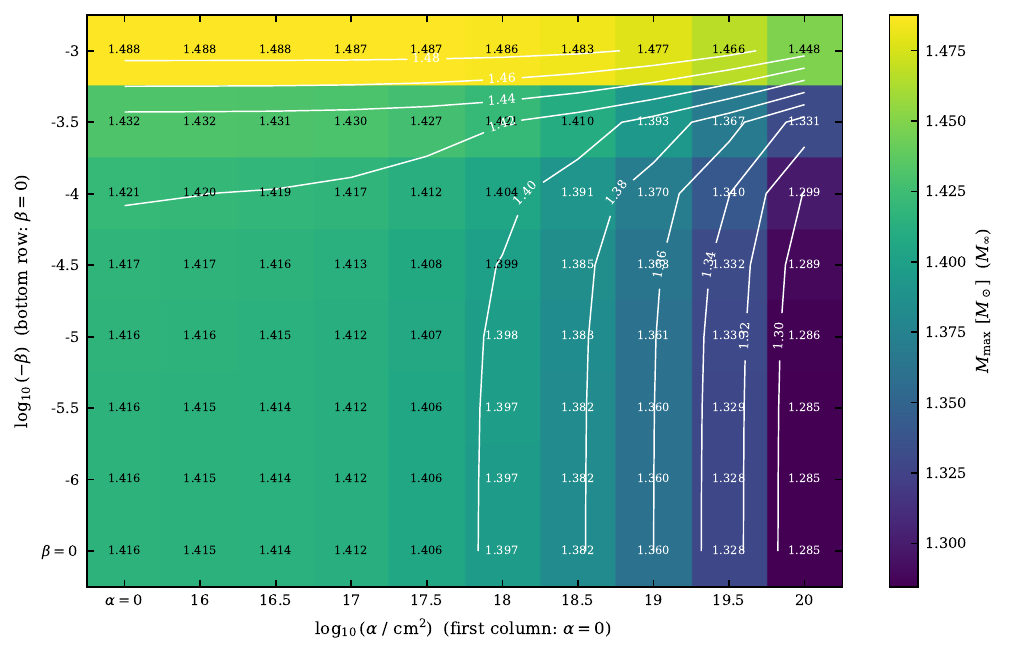}
\caption{Maximum mass $M_{\max}$ ($M_\infty$, in $M_\odot$) on the $(\alpha,\beta)$ plane (ideal equation of state). The first column corresponds to $\alpha=0$ and the bottom row to $\beta=0$. Contour lines connect equal values of $M_{\max}$; the numbers in the cells are listed in Table~\ref{tab:mmax_2d}.}
\label{fig:contour}
\end{figure*}

When both corrections are present, their effects on the maximum mass partly compensate each other. Table~\ref{tab:mmax_2d} lists $M_{\max}$ on the $(\alpha,\beta)$ plane. For $|\beta|\lesssim10^{-4}$ the two contributions add almost linearly. Different combinations of the couplings can therefore give the same maximum mass. For instance, $(\alpha,\beta)=(10^{17}\,\mathrm{cm^2},-7.8\times10^{-5})$, $(10^{18}\,\mathrm{cm^2},-2.5\times10^{-4})$ and $(10^{19}\,\mathrm{cm^2},-5.2\times10^{-4})$ all reproduce the GR value $1.4158\,M_\odot$. The maximum mass alone thus cannot identify the theory, and the full mass--radius relation is needed. We return to these combinations in Sec.~\ref{sec:joint}. For $\beta=-10^{-3}$ the trace coupling dominates, and the quadratic term is nearly irrelevant up to $\alpha\sim10^{18}\,\mathrm{cm^2}$, where it lowers the maximum mass by only $0.0016\,M_\odot$, compared with $0.019\,M_\odot$ for $\beta=0$. Figure~\ref{fig:contour} shows the same results as a map.
\begin{table*}[htbp]
\centering
\caption{Maximum mass $M_{\max}$ ($M_\infty$, in $M_\odot$) on the $(\alpha,\beta)$ grid. Columns give $\log_{10}(\alpha/{\rm cm}^2)$, with the first column corresponding to $\alpha=0$; rows give $\beta$.}
\label{tab:mmax_2d}
{\small\setlength{\tabcolsep}{4pt}
\begin{tabular}{lcccccccccc}
\toprule
$\beta$ & $\alpha=0$ & $16$ & $16.5$ & $17$ & $17.5$ & $18$ & $18.5$ & $19$ & $19.5$ & $20$ \\
\midrule
$0$ & 1.4158 & 1.4152 & 1.4141 & 1.4115 & 1.4063 & 1.3970 & 1.3822 & 1.3601 & 1.3283 & 1.2846 \\
$-10^{-6}$ & 1.4158 & 1.4153 & 1.4142 & 1.4116 & 1.4064 & 1.3971 & 1.3823 & 1.3602 & 1.3285 & 1.2847 \\
$-3.16\times10^{-6}$ & 1.4159 & 1.4154 & 1.4143 & 1.4117 & 1.4065 & 1.3973 & 1.3825 & 1.3604 & 1.3287 & 1.2850 \\
$-10^{-5}$ & 1.4163 & 1.4157 & 1.4146 & 1.4121 & 1.4069 & 1.3977 & 1.3831 & 1.3611 & 1.3295 & 1.2860 \\
$-3.16\times10^{-5}$ & 1.4173 & 1.4167 & 1.4157 & 1.4133 & 1.4082 & 1.3993 & 1.3849 & 1.3632 & 1.3321 & 1.2890 \\
$-10^{-4}$ & 1.4205 & 1.4201 & 1.4192 & 1.4171 & 1.4125 & 1.4042 & 1.3907 & 1.3701 & 1.3404 & 1.2989 \\
$-3.16\times10^{-4}$ & 1.4320 & 1.4318 & 1.4313 & 1.4299 & 1.4268 & 1.4206 & 1.4099 & 1.3929 & 1.3674 & 1.3311 \\
$-10^{-3}$ & 1.4876 & 1.4876 & 1.4876 & 1.4875 & 1.4871 & 1.4860 & 1.4833 & 1.4774 & 1.4665 & 1.4485 \\
\bottomrule
\end{tabular}}
\end{table*}

\subsection{Stability and energy conditions}
\label{sec:stability}
\begin{figure*}[htbp]
\centering
\includegraphics[width=\textwidth]{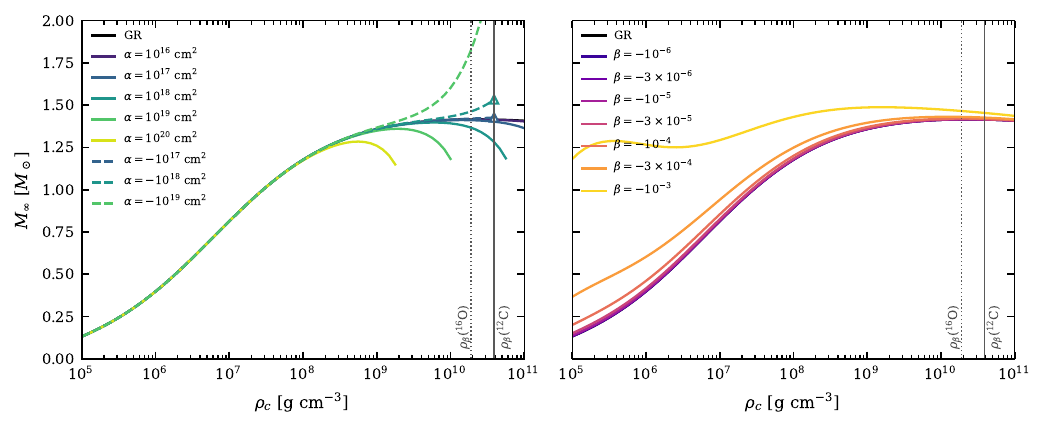}
\caption{Mass $M_\infty$ as a function of the central density. Left: the $\alpha$ family ($\beta=0$), including the $\alpha<0$ sequences of Fig.~\ref{fig:MR_alpha} (dashed). Right: the $\beta$ family ($\alpha=0$). Vertical lines mark the $^{12}$C (solid) and $^{16}$O (dotted) electron-capture densities. The maximum of each curve separates the stable (left) from the unstable (right) branch. For $\beta=-10^{-3}$ this applies to the high-density maximum only; the low-density behaviour depends on the surface density (Sec.~\ref{sec:stability}).}
\label{fig:Mrhoc}
\end{figure*}

Stability against radial perturbations can be read from the dependence of the mass on the central density, shown in Fig.~\ref{fig:Mrhoc}. For $\alpha\ge0$ each sequence has a single maximum. The configurations with $dM_\infty/d\rho_c>0$, to the left of the maximum, form the stable branch if the turning-point criterion \citep{HarrisonThorne1965} carries over, as it does for $\alpha=0$ (see below). The maximum-mass stars of Tables~\ref{tab:mmax_alpha} and \ref{tab:mmax_beta} then mark the onset of instability. Their central density decreases with increasing $\alpha$ or $|\beta|$, so the instability is reached in less compact, less relativistic stars. 

For $\alpha\le-10^{17}\,\mathrm{cm^2}$ there is no turning point below $\rho_\beta$, and the sequence ends for microphysical reasons. Whether these massive stars are dynamically stable cannot be settled by a standard radial-mode analysis. For a time-dependent rotated tetrad, the antisymmetric $tr$ component of Eq.~\eqref{eq:fieldeqs} no longer vanishes. It reads $(e^{B/2}-1)\,\partial_t f_T/r=0$ and, for $\alpha\neq0$, forbids any oscillation that changes $T$. A radial boost $\chi(t,r)$ must therefore be added to the tetrad, which leaves the metric unchanged. To linear order the same equation fixes it,
\begin{equation*}
    \chi\,e^{A/2}T'=(e^{B/2}-1)\,\partial_t\delta T.
\end{equation*}
Since $\delta T$ itself contains $\partial_t\chi$, one obtains $\delta T-\kappa\,\partial_t^2\delta T=\delta T_{\rm m}$, where $\delta T_{\rm m}$ is the contribution of the metric perturbations and $\kappa=4e^{-A}e^{-B/2}(e^{B/2}-1)^2/(rT')$. This is a local mode without spatial gradients. It is the linear manifestation of the extra Lorentz degree of freedom of $f(T)$ gravity \citep{Golovnev2021,Bahamonde2023}. Because $T$ has a minimum inside the star, $\kappa$ changes sign there, and the pulsation equations are not of Sturm--Liouville type. A consistent treatment of this mode is beyond the scope of this work. For the trace coupling alone ($\alpha=0$) no such mode arises. In that case the criterion $dM/d\rho_c>0$ has been confirmed by a mode analysis in $f(R,\mathcal{T})=R+2\lambda\mathcal{T}$ gravity \citep{Pretel2021}, as it has in Palatini $f(R)$ gravity \citep{Sarmah2022}. Figure~\ref{fig:Mrhoc} also shows that $\rho_\beta(^{16}\mathrm{O})$ lies below the GR turning point. For oxygen-rich cores, electron capture rather than the turning point limits the mass even in GR.  

The curve for $\beta=-10^{-3}$ in Fig.~\ref{fig:Mrhoc} behaves differently at low central density. Its mass remains above $1.18\,M_\odot$ down to $\rho_c=10^5\,\mathrm{g\,cm^{-3}}$, and it shows a second, low-density maximum. The reason is the effective pressure of the trace coupling,
\begin{equation*}
    |\beta|(\varepsilon-3p)/16\pi\simeq|\beta|\rho c^2/16\pi.
\end{equation*}
It grows linearly with density, like the pressure of an isothermal gas with sound speed $c\sqrt{|\beta|/16\pi}\approx1340\,\mathrm{km\,s^{-1}}$ for $\beta=-10^{-3}$. It dominates the degeneracy pressure wherever $c_s^2<|\beta|/16\pi$, i.e. below $\rho\approx2\times10^{5}\,\mathrm{g\,cm^{-3}}$. Configurations with $\rho_c\lesssim10^{6}\,\mathrm{g\,cm^{-3}}$ are therefore truncated isothermal spheres supported mainly by the trace coupling. Their mass is set by $\beta$ and by the surface density, not by $\rho_c$, as for a Bonnor--Ebert sphere. Indeed, at $\rho_c\simeq3\times10^{5}\,\mathrm{g\,cm^{-3}}$ the mass is $1.95$, $1.29$ and $0.46\,M_\odot$ for $\rho_s=10^{3}$, $10^{4}$ and $10^{5}\,\mathrm{g\,cm^{-3}}$, and the radius changes by a factor of five. The low-density maximum appears only for $\rho_s=10^{4}\,\mathrm{g\,cm^{-3}}$. The high-density maximum is unaffected: at $\rho_c=10^{9}\,\mathrm{g\,cm^{-3}}$ the three masses agree to within $5\times10^{-4}\,M_\odot$. Only this maximum has the usual meaning, and it is the one listed in Table~\ref{tab:mmax_beta}. The same effect explains the enhanced low-density masses for $\beta=-3\times10^{-4}$. Both values lie far outside the bound derived in Sec.~\ref{sec:obs}.

\begin{figure*}[htbp]
\centering
\includegraphics[width=\textwidth]{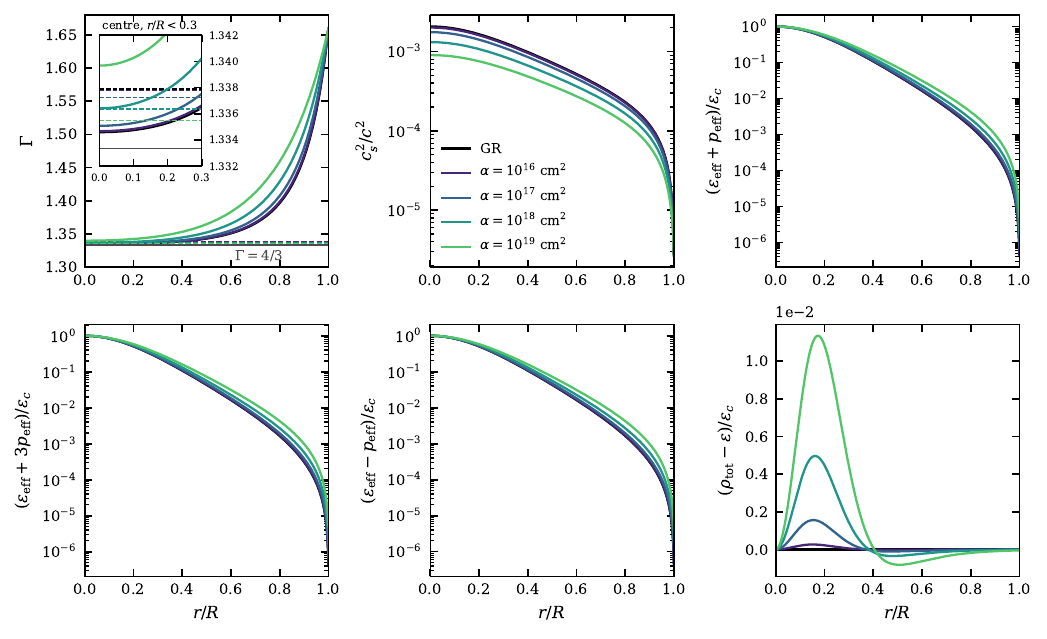}
\caption{Physical viability of the stars of Fig.~\ref{fig:struct_alpha}. Adiabatic index $\Gamma$ (dashed lines: the GR estimate $\frac43+1.12\,\frac{2GM}{Rc^2}$ of the critical value; inset: central region), squared sound speed, and the energy-condition combinations $\varepsilon_{\rm eff}+p_{\rm eff}$, $\varepsilon_{\rm eff}+3p_{\rm eff}$ and $\varepsilon_{\rm eff}-p_{\rm eff}$ normalized to $\varepsilon_c$. The last panel shows the difference between the total geometric energy density $\rho_{\rm tot}=-\mathring{G}^t{}_t/8\pi$ and $\varepsilon$.}
\label{fig:viab_alpha}
\end{figure*}

\begin{figure*}[htbp]
\centering
\includegraphics[width=\textwidth]{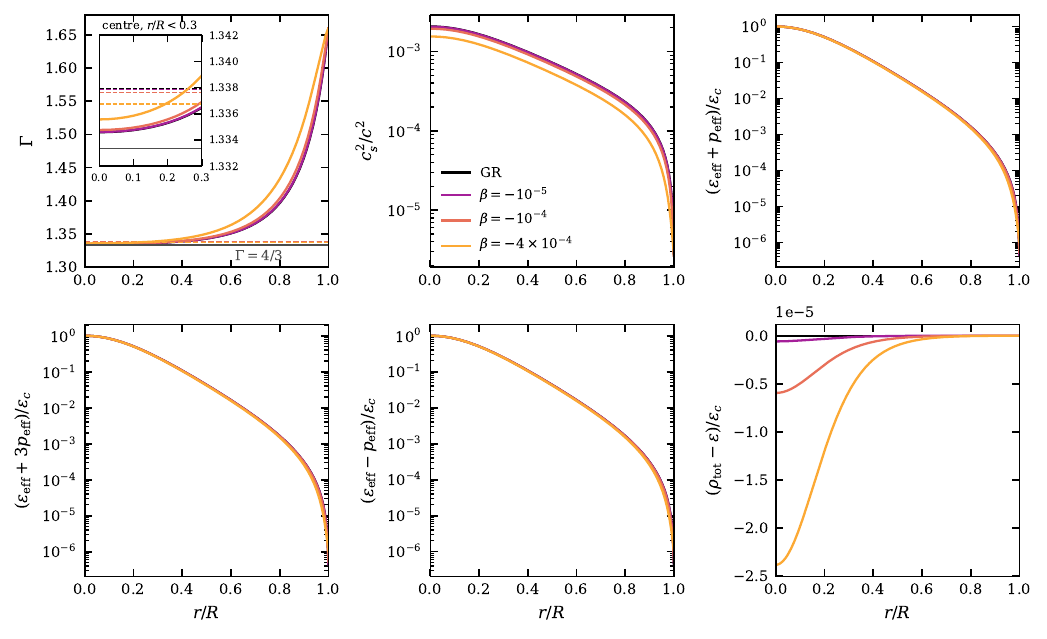}
\caption{As Fig.~\ref{fig:viab_alpha} for the stars of Fig.~\ref{fig:struct_beta}. For $\alpha=0$ the last panel equals $\beta(3\varepsilon-p)/16\pi\varepsilon_c$.}
\label{fig:viab_beta}
\end{figure*}

The adiabatic index $\Gamma=[(\varepsilon+p)/p]\,dp/d\varepsilon$, shown in the first panels of Figs.~\ref{fig:viab_alpha} and \ref{fig:viab_beta}, gives a complementary view. It rises from $\Gamma_c\simeq1.335$--$1.340$ at the centre, where the electrons are relativistic, to $5/3$ at the surface. The Newtonian condition $\Gamma>4/3$ is satisfied everywhere. In GR, however, the onset of instability is shifted to $\Gamma\simeq\frac43+K\,(2GM/Rc^2)$, with $K\approx1.12$ for a polytrope of index~3 \citep{Chandrasekhar1964}. For the present compactness this gives a critical value of $\simeq1.336$--$1.338$. The central values lie close to this threshold, as expected for maximum-mass stars. They increase slightly with $\alpha$ and $|\beta|$, consistent with the lower compactness. Since this estimate is derived within GR, we use it only as an indication and rely on the turning-point criterion above.

The remaining panels of Figs.~\ref{fig:viab_alpha} and \ref{fig:viab_beta} address causality and the energy conditions. The squared sound speed satisfies $c_s^2\le2.1\times10^{-3}$ throughout, so all configurations are causal. In terms of the effective fluid~\eqref{eq:effective}, the null, weak, strong and dominant energy conditions,
\begin{equation}
\varepsilon_{\rm eff}+p_{\rm eff}\ge0,\quad \varepsilon_{\rm eff}\ge0,\quad \varepsilon_{\rm eff}+3p_{\rm eff}\ge0,\quad \varepsilon_{\rm eff}-p_{\rm eff}\ge0,
\label{eq:ECs}
\end{equation}
are satisfied everywhere inside the stars, with $p_{\rm eff}/\varepsilon_{\rm eff}\le1.6\times10^{-3}$. The same holds for the combined stars, for example $(\alpha,\beta)=(10^{19}\,\mathrm{cm^2},-10^{-4})$. The last panels show the total energy density $\rho_{\rm tot}=-\mathring{G}^{t}{}_{t}/8\pi$, which also includes the geometric contribution of the $\alpha T^2$ term. For the $\alpha$ family this contribution is positive in the core, exceeding $\varepsilon$ by up to $1.1\%$ of $\varepsilon_c$. It becomes slightly negative in the outer layers ($r/R\gtrsim0.4$). Near the surface ($r/R\gtrsim0.99$) and outside the star, $\rho_{\rm tot}$ itself is negative. This small negative-energy region causes the decrease of $m(r)$ beyond the surface seen in Fig.~\ref{fig:exterior}. A similar sign change of the effective ``$f(T)$ fluid'' was found for polytropes in \citep{IlijicSossich2018}. The matter itself remains well behaved. The sign change is a property of the torsional correction rather than of the stellar matter. For the $\beta$ family, $\rho_{\rm tot}-\varepsilon=\beta(3\varepsilon-p)/16\pi$ is small and negative (Fig.~\ref{fig:viab_beta}). For all maximum-mass stars considered here $f_T>0.96$.

\subsection{Observational constraints}
\label{sec:obs}

\subsubsection{White dwarfs with dynamical masses: the bound on $\beta$}

The two couplings act on different parts of the mass--radius relation, so they are best confronted with different observations. The most reliable tests are white dwarfs whose masses are determined dynamically from binary orbits. These masses are independent of any assumed mass--radius relation. We use Sirius~B ($1.018\pm0.011\,M_\odot$, $R=5634\pm32\,\mathrm{km}$) \citep{Bond2017Sirius}, 40~Eri~B ($0.573\pm0.018\,M_\odot$, $R=9100\pm140\,\mathrm{km}$) \citep{Bond2017EriB} and Procyon~B ($0.592\pm0.006\,M_\odot$, $R=8571\pm223\,\mathrm{km}$) \citep{Bond2015Procyon}. They are shown in Figs.~\ref{fig:MR_alpha} and \ref{fig:MR_beta}.

A direct fit of the absolute radii is not meaningful with a zero-temperature equation of state. It neglects finite temperature, the hydrogen and helium envelopes and Coulomb corrections. Indeed, at $\rho_s=10^4\,\mathrm{g\,cm^{-3}}$ the GR model underestimates the radius of Sirius~B by $2.7\%$ and that of 40~Eri~B by $4.4\%$, while it agrees with Procyon~B. These effects shift the GR and modified-gravity radii in a similar way. We therefore use the radius shift at the observed mass, $\Delta R=R(M_{\rm obs};\alpha,\beta)-R_{\rm GR}(M_{\rm obs})$. We consider two exclusion criteria: (i) $|\Delta R|/R_{\rm GR}>3\%$, a systematic allowance for the neglected physics, and (ii) $|\Delta R|>2\sigma_{\rm tot}$, with $\sigma_{\rm tot}^2=\sigma_R^2+(0.025\,R_{\rm GR})^2$. Table~\ref{tab:dR} lists the shifts for $\alpha=0$. The radius shift grows rapidly with $|\beta|$. With $\rho_s=10^4\,\mathrm{g\,cm^{-3}}$, criterion (i) is crossed at $|\beta|=2.30\times10^{-5}$ and criterion (ii) at $|\beta|=4.43\times10^{-5}$. Both are set by 40~Eri~B. For $\beta=-10^{-4}$ the predicted radii exceed the GR ones by $9$--$15\%$, corresponding to $5$--$15\,\sigma_R$.
\begin{table*}[htbp]
\centering
\caption{Radius of the $\alpha=0$ model at the dynamical masses of Sirius~B, 40~Eri~B and Procyon~B ($\rho_s=10^4\,\mathrm{g\,cm^{-3}}$), and the shift $\Delta R=R_\beta-R_{\rm GR}$ in km and in units of the observational radius error $\sigma_R$. For $\beta=-10^{-3}$ no configuration with $\rho_c\geq10^5$\,g\,cm$^{-3}$ is lighter than $1.18\,M_\odot$.}
\label{tab:dR}
{\small\setlength{\tabcolsep}{4pt}
\begin{tabular}{lccccccccc}
\toprule
 & \multicolumn{3}{c}{Sirius B} & \multicolumn{3}{c}{40 Eri B} & \multicolumn{3}{c}{Procyon B} \\
\cmidrule(lr){2-4}\cmidrule(lr){5-7}\cmidrule(lr){8-10}
$\beta$ & $R$ [km] & $\Delta R$ [km] & $\Delta R/\sigma_R$ & $R$ [km] & $\Delta R$ [km] & $\Delta R/\sigma_R$ & $R$ [km] & $\Delta R$ [km] & $\Delta R/\sigma_R$ \\
\midrule
$0$ & 5480 & $+0$ & $+0.00$ & 8695 & $+0$ & $+0.00$ & 8546 & $+0$ & $+0.00$ \\
$-10^{-6}$ & 5485 & $+5$ & $+0.14$ & 8706 & $+11$ & $+0.08$ & 8556 & $+11$ & $+0.05$ \\
$-3\times10^{-6}$ & 5494 & $+14$ & $+0.42$ & 8729 & $+33$ & $+0.24$ & 8578 & $+32$ & $+0.14$ \\
$-10^{-5}$ & 5526 & $+45$ & $+1.42$ & 8807 & $+112$ & $+0.80$ & 8653 & $+107$ & $+0.48$ \\
$-3\times10^{-5}$ & 5618 & $+138$ & $+4.32$ & 9039 & $+344$ & $+2.45$ & 8875 & $+329$ & $+1.48$ \\
$-10^{-4}$ & 5968 & $+488$ & $+15.24$ & 9957 & $+1261$ & $+9.01$ & 9751 & $+1206$ & $+5.41$ \\
$-3\times10^{-4}$ & 7245 & $+1765$ & $+55.15$ & 14161 & $+5466$ & $+39.04$ & 13713 & $+5167$ & $+23.17$ \\
$-4\times10^{-4}$ & 8115 & $+2635$ & $+82.35$ & 18035 & $+9340$ & $+66.71$ & 17431 & $+8885$ & $+39.84$ \\
$-10^{-3}$ & -- & -- & -- & -- & -- & -- & -- & -- & -- \\
\bottomrule
\end{tabular}}
\end{table*}

\begin{figure*}[htbp]
\centering
\includegraphics[width=\textwidth]{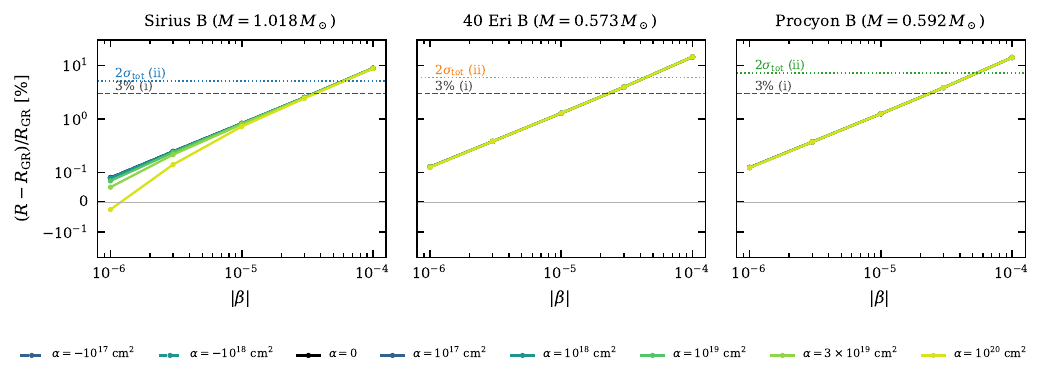}
\caption{Radius shift $(R(\alpha,\beta)-R_{\rm GR})/R_{\rm GR}$ at the dynamical masses of Sirius~B, 40~Eri~B and Procyon~B, against $|\beta|$, for $\alpha\in\{-10^{18},-10^{17},0,10^{17},10^{18},10^{19},3\times10^{19},10^{20}\}\,\mathrm{cm^2}$. Dashed: criterion (i), $3\%$. Dotted: criterion (ii), $2\sigma_{\rm tot}$. The $y$-axis is symmetric-logarithmic.}
\label{fig:dR}
\end{figure*}

\begin{figure*}[htbp]
\centering
\includegraphics[width=\textwidth]{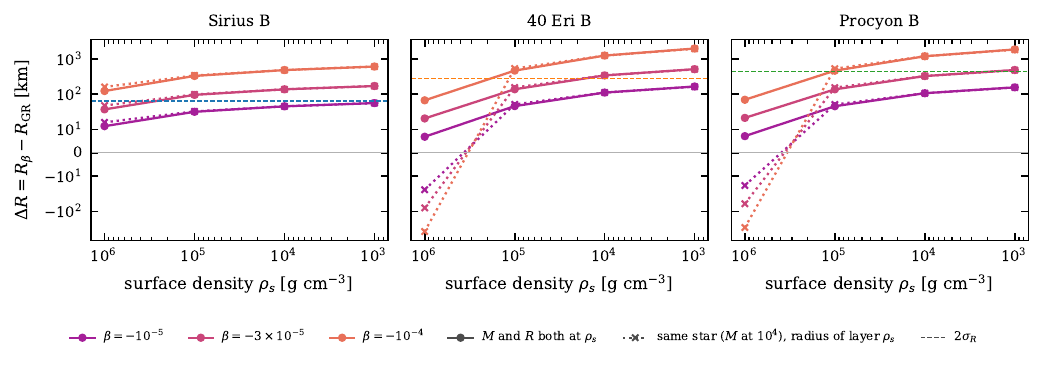}
\caption{Dependence of the radius shift $\Delta R=R_\beta-R_{\rm GR}$ on the surface density, at the observed masses ($\alpha=0$; $\beta=-10^{-5}$, $-3\times10^{-5}$, $-10^{-4}$). Solid lines with dots: the surface is placed at $\rho_s$, and both $M_\infty$ and $R$ are taken there. Dotted lines with crosses: the star is fixed by its mass (with the surface at $10^4\,\mathrm{g\,cm^{-3}}$), and $\Delta R$ is the shift of the layer where $\rho=\rho_s$; this shows where inside the star the inflation occurs. Negative values mean that the layer lies deeper than in GR. Dashed horizontal lines: twice the observed radius error, $2\sigma_R$, of each star.}
\label{fig:surface}
\end{figure*}

Figure~\ref{fig:dR} shows the same shifts for $\alpha$ between $-10^{18}$ and $10^{20}\,\mathrm{cm^2}$. The curves coincide where the criteria are crossed. At the observed masses, $\alpha$ changes the radii by at most $6\,\mathrm{km}$, while $\beta$ changes them by hundreds of km. The critical value of $\beta$ shifts by less than $2.3\times10^{-4}$ in $\log_{10}|\beta|$, i.e. by less than $0.05\%$, over the whole range of $\alpha$. The bound on $\beta$ is therefore independent of $\alpha$.
We now examine the robustness of this bound. Figure~\ref{fig:surface} shows that the radius shift caused by $\beta$ resides mostly in the low-density envelope. For 40~Eri~B and Procyon~B, $52$--$57\%$ of $\Delta R$ comes from layers with $\rho<10^5\,\mathrm{g\,cm^{-3}}$. At fixed mass, the layers with $\rho>10^6\,\mathrm{g\,cm^{-3}}$ even move inwards. This is expected from Eq.~\eqref{eq:dpdr}, since the term $|\beta|/c_s^2$ is largest where $c_s^2$ is smallest. Table~\ref{tab:bcrit} lists the bound for different surface densities. When the surface is moved inwards to $\rho_s=10^6\,\mathrm{g\,cm^{-3}}$, the bound loosens by a factor of about four, and Sirius~B takes over. We therefore quote two results:
\begin{equation}
|\beta|\lesssim(2.3\text{--}4.4)\times10^{-5}\quad(\rho_s=10^4\,\mathrm{g\,cm^{-3}}),\qquad
|\beta|\lesssim(1.0\text{--}1.8)\times10^{-4}\quad(\rho_s=10^6\,\mathrm{g\,cm^{-3}}).
\label{eq:betabound}
\end{equation}
The first assumes that the cold equation of state holds down to $10^4\,\mathrm{g\,cm^{-3}}$. The second uses only the degenerate interior and is conservative. In a real non-degenerate envelope the sound speed is much smaller than in degenerate matter. The term $|\beta|/c_s^2$ would then be even larger, so realistic envelopes are likely to strengthen the bound rather than weaken it. A quantitative statement requires envelope models with finite temperature, which we leave for future work.
\begin{table}[htbp]
\centering
\caption{Bound on $\beta$ ($\alpha=0$) from the three dynamical-mass white dwarfs for different surface densities $\rho_s$; both $M_\infty$ and $R$ are taken at $\rho_s$. Criterion (i): $|\Delta R|/R_{\rm GR}>3\%$. Criterion (ii): $|\Delta R|>2\sigma_{\rm tot}$. The star that sets each bound is given.}
\label{tab:bcrit}
\begin{tabular}{lcccc}
\toprule
$\rho_s$ [g\,cm$^{-3}$] & $\beta_{\rm crit}$ (i) & set by & $\beta_{\rm crit}$ (ii) & set by \\
\midrule
$10^{4}$ & $-2.30\times10^{-5}$ & 40 Eri B & $-4.43\times10^{-5}$ & 40 Eri B \\
$10^{5}$ & $-4.81\times10^{-5}$ & Procyon B$^a$ & $-8.14\times10^{-5}$ & Sirius B \\
$10^{6}$ & $-1.00\times10^{-4}$ & Sirius B & $-1.75\times10^{-4}$ & Sirius B \\
\bottomrule
\end{tabular}
\par\smallskip{\footnotesize $^a$The three stars agree within $0.5\%$.}
\end{table}

Through the correspondence $\lambda=\beta/2$, Eq.~\eqref{eq:betabound} also applies to $f(\mathring{R},\mathcal{T})=\mathring{R}+2\lambda\mathcal{T}$ gravity. It gives $|\lambda|\lesssim(1.2$--$2.2)\times10^{-5}$, or $|\lambda|\lesssim(0.5$--$0.9)\times10^{-4}$ conservatively. This improves the limit $\lambda\gtrsim-3\times10^{-4}$ obtained from massive white dwarfs \citep{Carvalho2017} by a factor of $14$--$26$, or $3$--$6$ conservatively. The improvement arises because the trace coupling acts most strongly on low-mass white dwarfs, for which precise dynamical masses and radii exist.

\subsubsection{Ultra-massive white dwarfs: the bound on $\alpha$}

The quadratic term cannot be tested in the same way. For $\alpha\le10^{20}\,\mathrm{cm^2}$ the mass--radius relation coincides with GR below $\sim1.2\,M_\odot$ (Fig.~\ref{fig:MR_alpha}). It can only be probed by ultra-massive white dwarfs. One of the most massive, ZTF~J1901+1458, has a radius of $2140^{+160}_{-230}\,\mathrm{km}$ and a mass of $1.327$--$1.365\,M_\odot$ \citep{Caiazzo2021,Desai:2025vbm}. Requiring that the maximum mass be at least $1.327\,M_\odot$ gives $\alpha\le3.30\times10^{19}\,\mathrm{cm^2}$ for $\beta=0$ (Table~\ref{tab:alphamax}). A negative $\beta$ raises the maximum mass and relaxes this bound slightly. Within the range $|\beta|\lesssim4.4\times10^{-5}$ allowed by Eq.~\eqref{eq:betabound}, we obtain
\begin{equation}
\alpha\lesssim(3.3\text{--}3.9)\times10^{19}\,\mathrm{cm^2}.
\label{eq:alphabound}
\end{equation}
We use only the mass of ZTF~J1901+1458. Its model-inferred radius is not used, because the comparison with it depends on the equation of state, as shown below. For comparison, the perihelion precession of Mercury gives $|\alpha|\lesssim1.1\times10^{30}\,\mathrm{cm^2}$ in our normalization \citep{DeBenedictisIlijic2016,Otoniel2026}, so white dwarfs are sensitive to couplings about ten orders of magnitude smaller. Neutron stars probe much smaller couplings, because the correction scales as $\alpha(GM/c^2)^2/R^4$ and is therefore much larger in more compact stars. Their bounds, however, depend strongly on the poorly known equation of state of dense matter. In \citep{LinChenZhai2022} the upper limit on $\alpha$ ranges from $-1.4\times10^{11}$ to $+4\times10^{10}\,\mathrm{cm^2}$, depending on the equation of state and on whether GW190814 is included; with GW190814 all equations of state require $\alpha<0$. The white-dwarf bound, in contrast, rests on the well-understood degenerate-electron equation of state and probes gravity at much lower densities.
\begin{table}[htbp]
\centering
\caption{Largest $\alpha$ for which $M_{\max}(\alpha,\beta)\geq1.327\,M_\odot$, the lower end of the model-inferred mass of ZTF~J1901+1458, with the central density and radius of the corresponding maximum-mass star.}
\label{tab:alphamax}
\begin{tabular}{lccc}
\toprule
$\beta$ & $\alpha_{\max}$ [cm$^2$] & $\rho_c$ [g\,cm$^{-3}$] & $R$ [km] \\
\midrule
$0$ & $3.30\times10^{19}$ & $1.02\times10^{9}$ & 2439 \\
$-10^{-6}$ & $3.31\times10^{19}$ & $1.02\times10^{9}$ & 2441 \\
$-3\times10^{-6}$ & $3.33\times10^{19}$ & $1.02\times10^{9}$ & 2445 \\
$-10^{-5}$ & $3.41\times10^{19}$ & $1.00\times10^{9}$ & 2459 \\
$-3\times10^{-5}$ & $3.68\times10^{19}$ & $9.51\times10^{8}$ & 2501 \\
$-10^{-4}$ & $4.78\times10^{19}$ & $7.99\times10^{8}$ & 2655 \\
$-3\times10^{-4}$ & $1.04\times10^{20}$ & $4.73\times10^{8}$ & 3181 \\
$-10^{-3}$ & $>10^{21}$ & -- & -- \\
\bottomrule
\end{tabular}
\end{table}

\begin{figure}[htbp]
\centering
\includegraphics[width=0.6\textwidth]{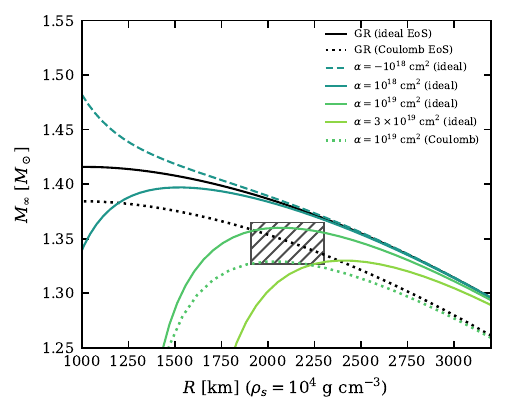}
\caption{Mass--radius relations near ZTF~J1901+1458 (hatched box: $M=1.327$--$1.365\,M_\odot$, $R=1910$--$2300\,\mathrm{km}$). Ideal equation of state: GR (solid black) and $\alpha=-10^{18}$ (dashed), $10^{18}$, $10^{19}$ and $3\times10^{19}\,\mathrm{cm^2}$ ($\beta=0$). Dotted: GR and $\alpha=10^{19}\,\mathrm{cm^2}$ with the Coulomb-corrected equation of state. The Coulomb correction lowers the curves by about $0.03\,M_\odot$ near the maximum; with it, the GR curve passes through the box.}
\label{fig:ZTF}
\end{figure}

The bound~\eqref{eq:alphabound} is indicative only. The mass of ZTF~J1901+1458 was inferred from a theoretical mass--radius relation rather than measured dynamically. The star is also hot, highly magnetized and rapidly rotating \citep{Caiazzo2021}, and none of these effects is included here. A stricter test would require the sequence to pass through the observed mass--radius box. Figure~\ref{fig:ZTF} shows that this test is not robust at present. With the ideal equation of state, the GR curve misses the box by $56\,\mathrm{km}$, while $\alpha\simeq10^{19}\,\mathrm{cm^2}$ passes through it. With the Coulomb-corrected equation of state, however, the GR curve itself passes through the box. The Coulomb correction lowers the maximum mass by about $2\%$, which is enough to change the outcome. We therefore do not use the box test as a constraint. For $\alpha<0$ the maximum mass at $\rho_\beta$ always exceeds $1.327\,M_\odot$, so this condition does not constrain negative $\alpha$. A dynamical mass for an ultra-massive white dwarf would turn Eq.~\eqref{eq:alphabound} into a robust bound.

\subsubsection{Joint constraints on $\alpha$ and $\beta$}
\label{sec:joint}

\begin{figure*}[htbp]
\centering
\includegraphics[width=\textwidth]{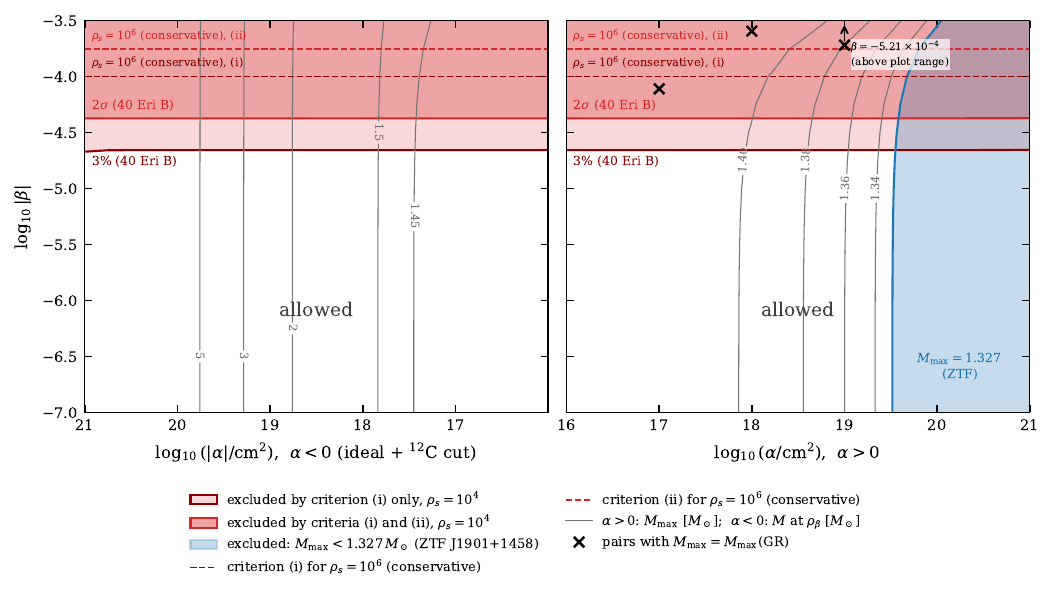}
\caption{Constraints in the $(\log_{10}|\alpha|,\log_{10}|\beta|)$ plane. Left: $\alpha<0$ (ideal equation of state, sequences stopped at $\rho_\beta(^{12}\mathrm{C})$). Right: $\alpha>0$. Shaded as excluded, with solid boundaries, for $\rho_s=10^4\,\mathrm{g\,cm^{-3}}$ and set by 40~Eri~B: light red, criterion (i) only ($2.30\times10^{-5}<|\beta|<4.43\times10^{-5}$); dark red, criteria (i) and (ii) ($|\beta|>4.43\times10^{-5}$). Blue: $M_{\max}<1.327\,M_\odot$ (ZTF~J1901+1458). Dashed: the conservative bounds for $\rho_s=10^6\,\mathrm{g\,cm^{-3}}$, $|\beta|=1.00\times10^{-4}$ (i) and $1.75\times10^{-4}$ (ii). Thin grey contours: $M_{\max}$ in $M_\odot$ for $\alpha>0$, and the mass at $\rho_\beta$ for $\alpha<0$. Crosses: the combinations of Fig.~\ref{fig:pairs} that reproduce the GR maximum mass; the one for $\alpha=10^{19}\,\mathrm{cm^2}$ ($\beta=-5.21\times10^{-4}$) lies above the plotted range and is marked by an arrow.}
\label{fig:allowed}
\end{figure*}

Figure~\ref{fig:allowed} combines the two constraints in the $(\alpha,\beta)$ plane. The radius criteria give horizontal boundaries, because the bound on $\beta$ does not depend on $\alpha$. The dashed lines show the conservative bounds for $\rho_s=10^6\,\mathrm{g\,cm^{-3}}$. The maximum-mass condition gives a nearly vertical boundary at $\alpha\simeq3.3\times10^{19}\,\mathrm{cm^2}$, which bends slightly towards larger $\alpha$ as $|\beta|$ grows. The thin contours show how the maximum mass varies across the plane. In the allowed region it stays between about $1.33$ and $1.42\,M_\odot$ for $\alpha>0$. For $\alpha<0$ only the bound on $\beta$ applies, and the mass at $\rho_\beta$ grows from about $1.43\,M_\odot$ to above $5\,M_\odot$ as $|\alpha|$ increases. The two couplings are thus constrained by different stars. This separation is the main observational result of this work.

\begin{figure*}[htbp]
\centering
\includegraphics[width=\textwidth]{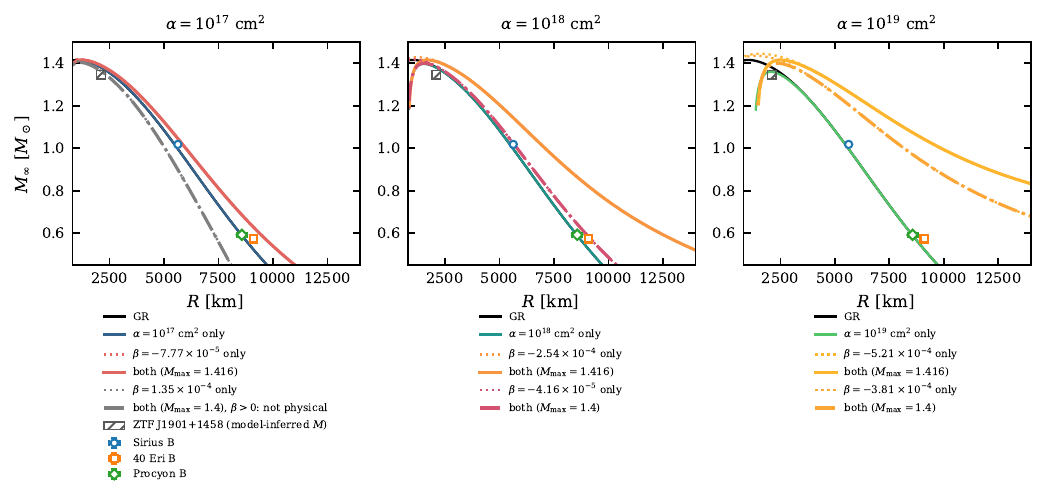}
\caption{Combinations of $\alpha$ and $\beta$ with the same maximum mass, for $\alpha=10^{17}$, $10^{18}$ and $10^{19}\,\mathrm{cm^2}$ (one panel each). Shown: GR (black), $\alpha$ only, $\beta$ only (dotted), and the combinations tuned to $M_{\max}=1.4158\,M_\odot$ (solid) and $1.40\,M_\odot$ (dash-dotted). The $\alpha=10^{17}\,\mathrm{cm^2}$, $1.40\,M_\odot$ combination requires $\beta>0$ and is not physical.}
\label{fig:pairs}
\end{figure*}

The separation also resolves the degeneracy in the maximum mass noted in Sec.~\ref{sec:MR}. Figure~\ref{fig:pairs} shows the combinations of $\alpha$ and $\beta$ that reproduce the GR maximum mass, or $M_{\max}=1.40\,M_\odot$. Below $1.3\,M_\odot$ each combination lies on the curve of its $\beta$-only partner. At $0.6\,M_\odot$ the GR-mimicking combinations for $\alpha=10^{17}$, $10^{18}$ and $10^{19}\,\mathrm{cm^2}$ have radii larger than in GR by $10.5\%$, $46\%$ and $160\%$. They all require $|\beta|\ge7.8\times10^{-5}$ and are excluded by the radius criteria at $\rho_s=10^4\,\mathrm{g\,cm^{-3}}$. The two with $\alpha\ge10^{18}\,\mathrm{cm^2}$ are excluded even by the conservative bound. Within the allowed range of $\beta$, the trace coupling shifts $M_{\max}$ by at most a few $\times10^{-3}\,M_\odot$. The combination for $\alpha=10^{17}\,\mathrm{cm^2}$ and $M_{\max}=1.40\,M_\odot$ would require $\beta>0$, which is not physical by Eq.~\eqref{eq:betarange}.

\section{Conclusions}
\label{sec:conclusions}

We have studied the equilibrium structure of white dwarfs in $f(T,\mathcal{T})$ gravity for the model $f=T+\alpha T^2+\beta\mathcal{T}$. The model combines a quadratic torsional correction with a coupling between geometry and the trace of the energy--momentum tensor. We formulated the theory in a single, consistent set of conventions. For the signature $(-,+,+,+)$ and $\mathcal{L}_m=p$, the trace coupling enters through $\beta(\Theta_{\mu\nu}-pg_{\mu\nu})+\frac{\beta}{2}\mathcal{T}g_{\mu\nu}$. Earlier compact-star studies in this theory used a combination that is inconsistent with this choice.

The single-coupling limits of the model are known theories. For $\alpha=0$ it is equivalent to $f(R,\mathcal{T})$ gravity, and for $\beta=0$ it is quadratic $f(T)$ gravity. Our solver reproduces published white-dwarf results in both limits \citep{Carvalho2017,Otoniel2026}. The trace coupling makes matter non-conserved and introduces a term $\propto\beta/c_s^2$ in the hydrostatic equation. Because the sound speed is very small in the outer layers of a white dwarf, regular stars require $\beta\le0$. Their outer layers are then extended, and the radius must be defined at a finite surface density. The quadratic term survives outside the star, so the mass must be taken at infinity. The difference from the surface mass, $M_\infty/m(R)-1\simeq2\alpha T(R)$, follows from the perturbative vacuum solution of \citep{DeBenedictisIlijic2016}. The mass difference exceeds $10^{-4}\,M_\odot$ for $\alpha\gtrsim10^{17}\,\mathrm{cm^2}$.

A positive $\alpha$ lowers the maximum mass, from $1.416\,M_\odot$ in GR to $1.285\,M_\odot$ at $\alpha=10^{20}\,\mathrm{cm^2}$, and moves it to lower central densities and larger radii. A negative $\alpha$ removes the turning point. The structure equations remain regular, and the mass is then limited by electron capture, reaching $1.53\,M_\odot$ at $\alpha=-10^{18}\,\mathrm{cm^2}$. A negative $\beta$ raises the maximum mass slightly. Its main effect, however, is to inflate low-mass white dwarfs. All stable configurations are causal. The effective matter satisfies the energy conditions, while the torsional correction carries a small negative energy density in the outer layers and outside the star.

The two couplings act on different parts of the mass--radius relation. White dwarfs with dynamical masses give $|\beta|\lesssim(2.3$--$4.4)\times10^{-5}$, independently of $\alpha$ for $-10^{18}\le\alpha\le10^{20}\,\mathrm{cm^2}$. This bound relies on the cold equation of state in the outer layers. Discarding layers below $10^6\,\mathrm{g\,cm^{-3}}$ gives the conservative bound $|\beta|\lesssim(1.0$--$1.8)\times10^{-4}$. Both improve the corresponding bound on $f(R,\mathcal{T})$ gravity. The ultra-massive white dwarf ZTF~J1901+1458 suggests $\alpha\lesssim(3.3$--$3.9)\times10^{19}\,\mathrm{cm^2}$. Combinations of $\alpha$ and $\beta$ that mimic the GR maximum mass are excluded by the radii of low-mass white dwarfs.

Several extensions would sharpen these results. The most important is a realistic envelope with finite temperature, which controls the robustness of the bound on $\beta$. A full equation of state with Coulomb corrections and composition effects would also reduce the systematic uncertainty in the radii. A radial-perturbation analysis for $\alpha\neq0$ requires a treatment of the extra Lorentz degree of freedom, which enters through a time-dependent boost of the tetrad. It is needed in particular for the massive stars with $\alpha<0$. Cooling and crystallization ages would probe the trace coupling in the non-degenerate envelope \citep{Vidal2025}. Rotation and magnetic fields are relevant for objects such as ZTF~J1901+1458. Finally, the gravitational redshift of white dwarfs in binaries offers an independent probe of the exterior mass correction.

\appendix

\section{Coulomb-corrected equation of state and comparison with quadratic $f(T)$ gravity}
\label{app:coulomb}

For the validation against \citep{Otoniel2026} we add the electrostatic energy of a body-centred cubic lattice of $^{12}$C ions \citep{Salpeter1961},
\begin{equation}
\varepsilon_L = C_{\rm bcc}\,e^2\,n_e^{4/3}Z^{2/3},\qquad p_L=\frac{\varepsilon_L}{3},\qquad C_{\rm bcc}=-1.444231,
\label{eq:coulomb}
\end{equation}
with $Z=6$, so that $\varepsilon=\rho c^2+\varepsilon_e+\varepsilon_L$ and $p=p_e+p_L$. The pressure remains positive above $12.8\,\mathrm{g\,cm^{-3}}$, and $c_s^2>0$ above $6.6\,\mathrm{g\,cm^{-3}}$. Both conditions therefore hold for all $\rho\ge\rho_s$. The lattice reduces the pressure by $11\%$ at $10^4\,\mathrm{g\,cm^{-3}}$ and by $1.5\%$ at $10^{10}\,\mathrm{g\,cm^{-3}}$. For this equation of state, equality of the Gibbs energies of the parent and daughter lattices gives $\rho_\beta(^{12}\mathrm{C})=4.157\times10^{10}\,\mathrm{g\,cm^{-3}}$, in agreement with the value $4.158\times10^{10}\,\mathrm{g\,cm^{-3}}$ of \citep{Otoniel2026}.

\begin{figure*}[htbp]
\centering
\includegraphics[width=\textwidth]{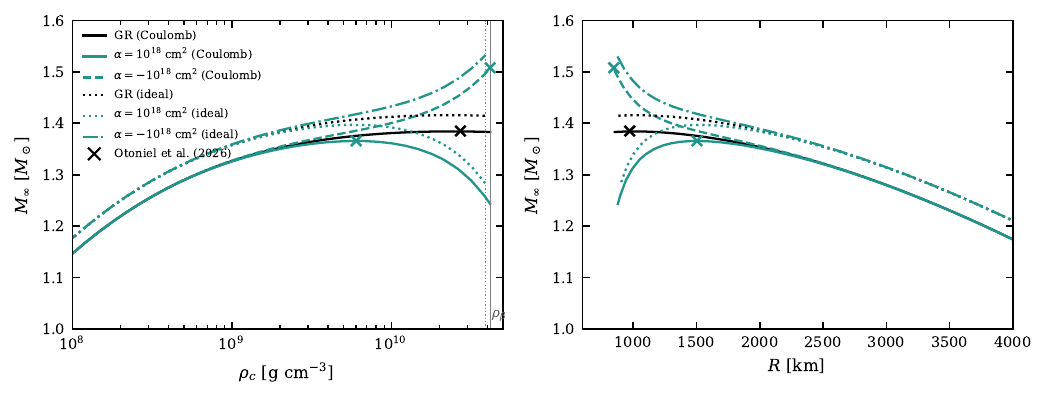}
\caption{Comparison with quadratic $f(T)$ gravity \citep{Otoniel2026}. $M_\infty$ against $\rho_c$ (left) and $R$ (right) for GR and $\alpha=\pm10^{18}\,\mathrm{cm^2}$ ($\beta=0$). Solid and dashed: Coulomb-corrected equation of state with the electron-capture cut. Dotted and dash-dotted: ideal equation of state. Crosses: the values of \citep{Otoniel2026} for the GR maximum, the first maximum for $\alpha=10^{18}\,\mathrm{cm^2}$, and the star at $\rho_\beta$ for $\alpha=-10^{18}\,\mathrm{cm^2}$ ($1.508\,M_\odot$, $846.8\,\mathrm{km}$).}
\label{fig:coulomb}
\end{figure*}

Figure~\ref{fig:coulomb} compares the two equations of state for $\alpha=0$ and $\pm10^{18}\,\mathrm{cm^2}$. The lattice correction lowers the maximum mass by $2.2\%$. The comparison with \citep{Otoniel2026} is summarized in Table~\ref{tab:validation}.

\section{Surface definition for $\beta<0$}
\label{app:surface}

\begin{figure*}[htbp]
\centering
\includegraphics[width=\textwidth]{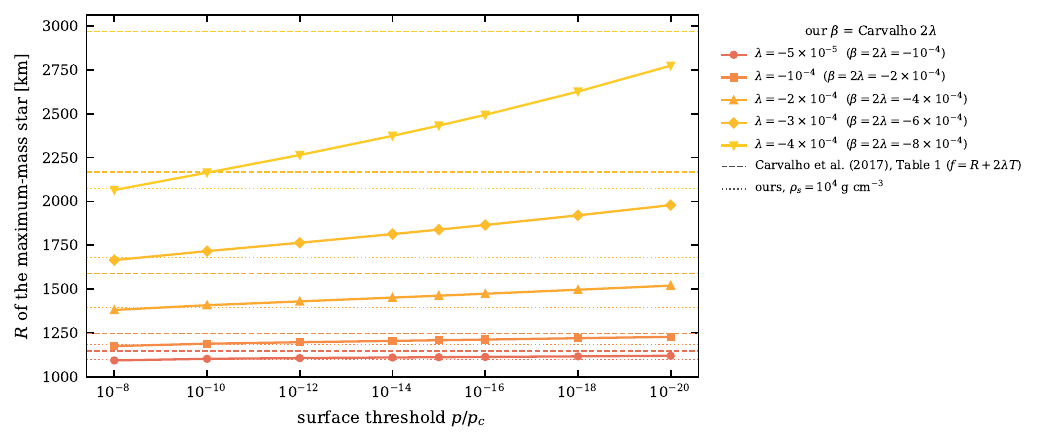}
\caption{Radius of our maximum-mass star, at our central density, against the surface threshold $p/p_c$ ($\alpha=0$), for the five cases of Table~1 of \citep{Carvalho2017} ($f=R+2\lambda\mathcal{T}$, so that $\beta=2\lambda$). Dashed: their radii, at their central densities. Dotted: our radii at $\rho_s=10^4\,\mathrm{g\,cm^{-3}}$.}
\label{fig:carvalho}
\end{figure*}

For $\beta<0$ the effective pressure $p_{\rm eff}=p+|\beta|(\varepsilon-3p)/16\pi$ does not vanish where $p\to0$. The envelope is then supported by $|\beta|\varepsilon/16\pi$ and resembles an isothermal atmosphere. A radius defined by a pressure threshold therefore grows without limit as the threshold is lowered. Figure~\ref{fig:carvalho} shows this for the five cases of Table~1 of \citep{Carvalho2017}. At $\lambda=-4\times10^{-4}$ the radius grows from $2065\,\mathrm{km}$ at $p/p_c=10^{-8}$ to $2775\,\mathrm{km}$ at $10^{-20}$, and is still rising. The curves in Fig.~\ref{fig:carvalho} are for our maximum-mass stars. Since the maxima are very flat, our central densities differ from those of \citep{Carvalho2017} by $1$--$8\%$, for example $3.09\times10^{9}$ against $3.37\times10^{9}\,\mathrm{g\,cm^{-3}}$ for $\lambda=-4\times10^{-4}$ (Table~\ref{tab:validation}). For a direct comparison of radii we therefore also computed stars at their central densities. At $p/p_c=10^{-20}$ these agree with their radii within $0.8\%$ for $\lambda=-0.5\times10^{-4}$ and $2\%$ for $\lambda=-10^{-4}$. For larger $|\lambda|$ their ``$p=0$'' surface lies even further out. The masses are insensitive to the surface definition and agree within $10^{-3}\,M_\odot$. Radii quoted for a $p=0$ surface in trace-coupled theories with $\lambda<0$ should therefore be interpreted with care. This is why we define the radius at a fixed rest-mass density.

\section*{Data availability}
All results of this work follow from the equations given in the article, and the data needed to reproduce them are contained in its tables and figures. The observational data are taken from \citet{Bond2017Sirius,Bond2017EriB,Bond2015Procyon,Caiazzo2021}.
\section*{Acknowledgments}
R.S. acknowledges the hospitality and research facilities of the Institute of Physics, Bhubaneswar, India, during her visit.

\begin{contribution}
 
Both authors contributed equally to this work.


\end{contribution}

\bibliography{sample701}{}
\bibliographystyle{aasjournalv7}



\end{document}